\documentstyle[12pt]{article}
\newlength{\pubnumber} 

\catcode`\@=11
\@addtoreset{equation}{section}

\def\section{\@startsection{section}{1}{\z@}{3.5ex plus 1ex minus .2ex}
 {2.3ex plus .2ex}{\large\bf}}
\def\subsection{\@startsection{subsection}{2}{\z@}{2.3ex plus .2ex}
 {2.3ex plus .2ex}{\bf}}

\begin{document}

\begin{titlepage}
\samepage{
\rightline{CERN-TH/98-370}
\rightline{\tt hep-ph/9811428}
\rightline{November 1998}
\vfill
\begin{center}
  {\Large \bf Light Neutrinos without Heavy Mass Scales:\\
           A Higher-Dimensional Seesaw Mechanism\\}
\vfill
\medskip
   {\large
    Keith R. Dienes,$^1$
    Emilian Dudas,$^{1,2}$
     $\,$and$\,$
      Tony Gherghetta$^1$\footnote{
     E-mail addresses:  ~keith.dienes, $\,$emilian.dudas,
    $\,$tony.gherghetta@cern.ch}
    \\}
\vspace{.18in}
 {\it  $^1$ CERN Theory Division, CH-1211 Geneva 23, Switzerland\\}
\vspace{.04in}
 {\it  $^2$ LPTHE, Univ.\ Paris-Sud, F-91405, Orsay Cedex, France\footnote{
         Laboratoire associ\'e au CNRS-URA-D0063.}\\}
\end{center}
\vfill
\begin{abstract}
  {\rm
     Recent theoretical developments have shown that extra spacetime
      dimensions can lower the fundamental GUT, Planck, and string scales.
     However, recent evidence for neutrino oscillations suggests the
       existence of light non-zero neutrino masses, which in turn suggests
     the need for a heavy mass scale via the seesaw mechanism.
     In this paper, we make several observations in this regard.
     First, we point out that allowing the right-handed neutrino
     to experience extra spacetime dimensions naturally permits the left-handed
     neutrino mass to be power-law suppressed relative to the masses
     of the other fermions.   This occurs due to the power-law running
      of the neutrino Yukawa couplings, and therefore does not require
      a heavy scale for the right-handed neutrino.  Second, we show
      that a higher-dimensional analogue of the seesaw mechanism may also
     be capable of generating naturally
     light neutrino masses without the introduction of a heavy mass scale.
      Third, we show that such a higher-dimensional
     seesaw mechanism may even be able to explain neutrino oscillations
      {\it without}\/ neutrino masses, with oscillations induced indirectly
         via the masses of the Kaluza-Klein states.
     Fourth, we point out that even when higher-dimensional right-handed
       neutrinos are given a bare Majorana mass, the higher-dimensional
        seesaw mechanism surprisingly replaces this mass scale with
         the radius scale of the extra dimensions.  Finally, we also
        discuss a possible new mechanism for
       inducing lepton-number violation
       by shifting the positions of D-branes in Type~I string theory.}
\end{abstract}
\smallskip}
\end{titlepage}

\setcounter{footnote}{0}

\newcommand{\newc}{\newcommand}

\newc{\gsim}{\lower.7ex\hbox{$\;\stackrel{\textstyle>}{\sim}\;$}}
\newc{\lsim}{\lower.7ex\hbox{$\;\stackrel{\textstyle<}{\sim}\;$}}

\def\beq{\begin{equation}}
\def\eeq{\end{equation}}
\def\beqn{\begin{eqnarray}}
\def\eeqn{\end{eqnarray}}
\def\dnot#1{{ \not{\!\! {#1}} }}  
\def\eps{{\epsilon}}
\def\nn{{\bf n}}
\def\sosixteen{{$SO(16)\times SO(16)$}}
\def\e8{{$E_8\times E_8$}}
\def\V#1{{\bf V_{#1}}}
\def\half{{\textstyle{1\over 2}}}
\def\ttwo{{\vartheta_2}}
\def\tthree{{\vartheta_3}}
\def\tfour{{\vartheta_4}}
\def\ttwob{{\overline{\vartheta}_2}}
\def\tthreeb{{\overline{\vartheta}_3}}
\def\tfourb{{\overline{\vartheta}_4}}
\def\etainv{{\overline{\eta}}}
\def\Str{{{\rm Str}\,}}
\def\bone{{\bf 1}}
\def\chibar{{\overline{\chi}}}
\def\Jbar{{\overline{J}}}
\def\qbar{{\overline{q}}}
\def\calO{{\cal O}}
\def\calE{{\cal E}}
\def\calT{{\cal T}}
\def\calM{{\cal M}}
\def\calN{{\cal N}}
\def\calF{{\cal F}}
\def\calY{{\cal Y}}
\def\rep#1{{\bf {#1}}}
\def\ie{{\it i.e.}\/}
\def\eg{{\it e.g.}\/}
\def\eleven{{(11)}}
\def\ten{{(10)}}
\def\nine{{(9)}}
\def\Ip{{\rm I'}}
\def\oneprime{{I$'$}}
\hyphenation{su-per-sym-met-ric non-su-per-sym-met-ric}
\hyphenation{space-time-super-sym-met-ric}
\hyphenation{mod-u-lar mod-u-lar--in-var-i-ant}


\def\inbar{\,\vrule height1.5ex width.4pt depth0pt}

\def\IC{\relax\hbox{$\inbar\kern-.3em{\rm C}$}}
\def\IQ{\relax\hbox{$\inbar\kern-.3em{\rm Q}$}}
\def\IR{\relax{\rm I\kern-.18em R}}
 \font\cmss=cmss10 \font\cmsss=cmss10 at 7pt
\def\IZ{\relax\ifmmode\mathchoice
 {\hbox{\cmss Z\kern-.4em Z}}{\hbox{\cmss Z\kern-.4em Z}}
 {\lower.9pt\hbox{\cmsss Z\kern-.4em Z}}
 {\lower1.2pt\hbox{\cmsss Z\kern-.4em Z}}\else{\cmss Z\kern-.4em Z}\fi}

\def\NPB#1#2#3{{\it Nucl.\ Phys.}\/ {\bf B#1} (19#2) #3}
\def\PLB#1#2#3{{\it Phys.\ Lett.}\/ {\bf B#1} (19#2) #3}
\def\PRD#1#2#3{{\it Phys.\ Rev.}\/ {\bf D#1} (19#2) #3}
\def\PRL#1#2#3{{\it Phys.\ Rev.\ Lett.}\/ {\bf #1} (19#2) #3}
\def\PRT#1#2#3{{\it Phys.\ Rep.}\/ {\bf#1} (19#2) #3}
\def\CMP#1#2#3{{\it Commun.\ Math.\ Phys.}\/ {\bf#1} (19#2) #3}
\def\MODA#1#2#3{{\it Mod.\ Phys.\ Lett.}\/ {\bf A#1} (19#2) #3}
\def\IJMP#1#2#3{{\it Int.\ J.\ Mod.\ Phys.}\/ {\bf A#1} (19#2) #3}
\def\NUVC#1#2#3{{\it Nuovo Cimento}\/ {\bf #1A} (#2) #3}
\def\etal{{\it et al.\/}}

\long\def\@caption#1[#2]#3{\par\addcontentsline{\csname
  ext@#1\endcsname}{#1}{\protect\numberline{\csname
  the#1\endcsname}{\ignorespaces #2}}\begingroup
    \small
    \@parboxrestore
    \@makecaption{\csname fnum@#1\endcsname}{\ignorespaces #3}\par
  \endgroup}
\catcode`@=12

\input epsf


\section{Introduction}
\setcounter{footnote}{0}

Recent theoretical developments have shown
that extra spacetime dimensions have the potential to lower the
fundamental GUT scale~\cite{DDG},
the fundamental Planck scale~\cite{Dim}, and the fundamental string
scale~\cite{Witten,lykken,Dim,henry,DDG,bachas}.
The extra dimensions that lower the GUT scale are ``universal'',
and are felt by all forces, both gauge and gravitational.
Those that lower the Planck scale, by contrast, are felt only
by the gravitational interaction.  Together, both types of
extra dimensions can conspire to lower the string scale.
Indeed, by imagining extra spacetime dimensions
of various types and sizes,
it may even be possible to lower all of these scales to the
TeV range, although this is probably only an interesting (and likely
unrealistic) extrapolation.  However, the important lesson from
these developments is that the fundamental high energy scales of
physics are not immutable, and that taking extra spacetime dimensions
seriously as physical entities permits these energy scales to
be lower (perhaps even
substantially lower) than they are typically imagined to be on the basis
of four-dimensional extrapolations from low-energy data.
More recently, implications of these ideas have been considered in
cosmology~\cite{blackholes,DDGR,AHDMR,sacha,lyth},
in radius stabilization~\cite{DDGR,AHDMR},
and even in potentially explaining the
fermion mass hierarchy~\cite{DDG,Ross} and the properties
of soft SUSY-breaking parameters~\cite{softmasses,sundrum}.
Earlier discussions of TeV-scale extra dimensions can also be
found in Ref.~\cite{Antoniadis}, and
possible collider signatures of such extra dimensions
are discussed in Refs.~\cite{Antoniadis,colliders}.
General consequences of this new ``brane world''
picture of extra spacetime dimensions
and reduced energy scales are also discussed in Ref.~\cite{braneworld}.

At first glance, this situation may seem to suggest that there is no
further need for high energy scales.
However, as has recently been emphasized in Refs.~\cite{pati,ramond,wilczek},
low-energy neutrino data provide independent evidence for yet another
high mass scale.
Specifically, if neutrinos have light but non-zero masses (as suggested
by recent SuperKamiokande data~\cite{superK}), then these masses are most
naturally
explained in the context of $SO(10)$ unification
via the seesaw mechanism~\cite{seesaw}.
However, the seesaw mechanism relies on the existence of a new
heavy mass scale $M$ associated with a right-handed neutrino singlet field $N$.
Indeed, in the simplest scenarios, light neutrino masses
in the $10^{-2}$ eV range imply that
$M$ should be of the
same order of magnitude as the usual four-dimensional GUT
scale $\approx 10^{16}$ GeV.  This therefore
provides a further need for a high fundamental
GUT scale.  In Ref.~\cite{wilczek}, this is referred to as
``the third pillar of unification'', and we agree that this observation
should not be taken lightly.

In this paper, we shall therefore consider how light neutrino masses may
be generated without the introduction of heavy mass scales.
Our goal is to suggest a number
of higher-dimensional mechanisms which might permit naturally light neutrino
masses to be generated.
Our starting point is the observation that
because the right-handed neutrino is a Standard-Model gauge singlet,
it need not be restricted to a ``brane'' with respect to the
full higher-dimensional space.
It is therefore possible for this field
to experience extra spacetime dimensions and thereby
accrue an infinite tower of Kaluza-Klein excitations.
This then leads to a number of higher-dimensional mechanisms
for suppressing the resulting neutrino mass without a heavy mass scale,
and in this paper we shall make five specific observations.

\begin{itemize}

\item  First, we point out that allowing the right-handed
    neutrino to feel extra spacetime dimensions naturally permits the
    resulting left-handed neutrino mass to be power-law suppressed relative
    to the masses of all of the other Standard-Model fermions.  This
    occurs due to the power-law running of the neutrino Yukawa couplings,
    and can therefore drive the neutrino Yukawa couplings to extremely small
    values over a very short energy interval.

\item  Second, if the right-handed neutrino has a corresponding tower of
     Kaluza-Klein excitations, then the usual seesaw mechanism must
      be generalized to reflect mixings between
      the left-handed neutrino and the full tower of
      right-handed Kaluza-Klein states.  We therefore examine
      some of the consequences of such a higher-dimensional
       seesaw mechanism, and show that such a higher-dimensional
       seesaw mechanism may also be capable of generating a naturally
     light neutrino Majorana mass without the introduction of a heavy
right-handed
      neutrino mass scale.

\item Third, we show that
       a higher-dimensional seesaw mechanism may even be able to
      explain neutrino oscillations {\it without}\/
      neutrino masses, with oscillations induced indirectly via the
       masses of the Kaluza-Klein states.  This would therefore represent
       a radical departure from the usual four-dimensional situation in
       which neutrino oscillations are taken as evidence for
       neutrino masses.

\item Fourth, we point out that even when the right-handed neutrino
      is given a bare Majorana mass, our higher-dimensional seesaw
      mechanism essentially replaces this mass scale with the
      radius scale of the extra spacetime dimensions.  This replacement
      arises due to the cumulative effects of the Kaluza-Klein states,
      and is therefore
      also surprising from a na\"\i ve four-dimensional point of view.

\item  Finally, one of the crucial questions in explaining neutrino
       masses and oscillations is the violation of lepton number.  We
      therefore propose, within the context of our higher-dimensional
      seesaw mechanism, a new mechanism for generating lepton-number
     violation.  This method involves shifting the positions
      of D-branes in Type~I string theory.

\end{itemize}

These five mechanisms all have, as their basic goal, the generation
of light neutrino masses without the use of a heavy mass scale.
We therefore propose
these mechanisms in the expectation that they
are likely to play an important role in
any future systematic analysis of neutrino masses in theories with large
extra spacetime dimensions.

\section{Higher-dimensional mechanisms for light neutrino masses}
\setcounter{footnote}{0}

\subsection{Review:  ~The usual seesaw mechanism}

Let us begin by briefly reviewing the usual $SO(10)$
seesaw mechanism~\cite{seesaw}.
We imagine that there exists a right-handed neutrino (henceforth denoted
$N$), and that there exists a set of mass terms for $\calN\equiv (\nu_L,N)$
of the form ${\cal N} \calM {\cal N}^T$, where
\beq
            \calM ~=~ \pmatrix{ 0 & m \cr
                     m & M \cr}~.
\label{usualmatrix}
\eeq
Note that for the purposes of this discussion, we shall ignore
possible non-diagonality in flavor indices.
In this matrix, the entry $m$ arises as a standard Yukawa coupling
resulting from electroweak symmetry breaking,
\beq
          m ~\approx~ y_\nu \langle \phi \rangle~,
\label{Higgsvev}
\eeq
where $\langle \phi\rangle\approx 246$ GeV is the electroweak
Higgs VEV.
Since the neutrino Yukawa coupling $y_\nu$ is presumed
(on the basis of naturalness arguments) to be of order one,
we expect $m \approx {\cal O}$($10^2$ GeV).
The entry $M$, by contrast, is a Majorana mass for the right-handed
singlet $N$, and is presumed to arise through the breaking of
the $SO(10)$ GUT symmetry.  Such a term can arise, for example,
through the use of a large $SO(10)$ representation such as the
${\bf 126}$ representation (the five-index totally antisymmetric
tensor).  Thus, in the usual scenario, we expect that
$M\approx 10^{16}$ GeV.
By diagonalizing the mass matrix (\ref{usualmatrix}), we then obtain
the two mass eigenvalues
\beq
         \lambda_{\pm} ~=~ \half\left( M \pm \sqrt{M^2+4m^2}\right)~~~~~
         \Longrightarrow~~~~
               \lambda_- ~\approx~ - {m^2\over M}~,~~~~~
               \lambda_+ ~\approx~ M~
\label{usualseesaw}
\eeq
to leading order in $m/M$.
The physical light neutrino state is then interpreted as the
linear combination corresponding to the mass eigenvalue $\lambda_-$,
with mass $|\lambda_-|$.
Thus, the presence of the heavy mass scale $M$ serves to suppress
the neutrino mass so that it comes out substantially below the electroweak
scale.

This scenario is very simple and elegant.  In the context of string theory,
however, certain difficulties may arise.
The most pressing of these concerns the generation of the required
Majorana mass $M$ for the right-handed neutrino $N$.
As we remarked above, this is typically achieved in field theory through
the use of a {\bf 126} representation.  However, within the context
of a wide class of string $SO(10)$ GUT models, it has
been shown~\cite{no126,review}
that {\bf 126} representations generically do not arise.  Other possibilities
include simulating the effects of {\bf 126} representations through
tensor products of smaller representations~\cite{babu}, but even this
has been shown to be difficult within the context of string GUT
models~\cite{no126}.

\subsection{A higher-dimensional seesaw mechanism:  General setup}

Let us now consider how we might generate suppressed neutrino masses
 {\it without}\/ the introduction of such a high mass scale.
As we discussed in the Introduction, our goal
is to lay out a number
of alternatives within the context of theories with extra large spacetime
dimensions.  To this end, the first thing we notice is that unlike all of
the other Standard-Model fermions, the right-handed
neutrino $N$ is a Standard-Model singlet.  Thus,
this field need not necessarily be restricted to the ``brane'' that contains
the
remaining  Standard-Model fermions --- \ie, it is possible that this field
experiences
extra spacetime dimensions and thereby accrues an infinite
tower of Kaluza-Klein excitations.
For simplicity, we shall assume the appearance of one extra spacetime
dimension of radius $R$, so that the mass of the $n^{\rm th}$ Kaluza-Klein
state $N^{(n)}$ is given by
\beq
               m_n ~\approx~ {n/R}~,~~~~n\in\IZ~.
\label{KKmasses}
\eeq
For the purposes of this qualitative discussion, it
will not be necessary to specify whether this extra dimension
is ``universal'' (\ie, experienced by the Standard-Model gauge
bosons and Higgs fields as well as by gravity), or only
gravitational.
Therefore we shall not need to specify whether $R^{-1}\gsim {\cal O}$(TeV),
as is required in the first case, or $R^{-1}\lsim {\cal O}$(TeV),
as permitted in the second case.
In either case,
the important point is that $R^{-1}$ may be taken to be substantially
below the usual four-dimensional GUT, Planck, or string scales.
We also note that the following discussion continues to hold if
more than one extra dimension are considered.

\begin{figure}[ht]
\centerline{ \epsfxsize 3.5 truein \epsfbox {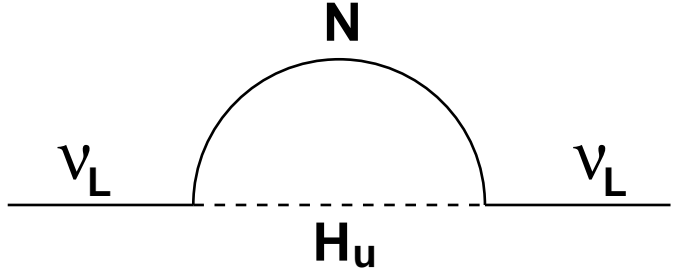}}
\caption{Typical one-loop diagram that can induce power-law running
   of the neutrino Yukawa coupling as a result of Kaluza-Klein
   states for the right-handed neutrino field $N$.  If only the
   right-handed neutrino $N$ experiences the extra dimensions,
   then the Yukawa coupling for the neutrino can be power-law
   suppressed relative to the Yukawa couplings for all other
   matter fields.}
\label{thefigure}
\end{figure}

There are two immediate consequences of introducing a Kaluza-Klein tower
for $N$.
The first, as discussed in Ref.~\cite{DDG},
is the power-law running that this induces for the Yukawa coupling $y_\nu$
through diagrams such as shown in Fig.~\ref{thefigure}.
In such diagrams, the presence of an infinite tower of Kaluza-Klein states
in the loop causes the evolution of the Yukawa coupling to accrue a
power-law behavior which can drive the Yukawa coupling $y_\nu$ to extremely
small values over a very short energy internal.
Thus, we see that
a Kaluza-Klein tower for the right-handed neutrino provides a natural
way of suppressing the value of the Yukawa coupling $y_\nu$ and thereby
suppressing $m$.
Detailed calculations of the resulting neutrino masses would then proceed
along the lines discussed in Refs.~\cite{DDG,Ross}.

The second observation\footnote{This possibility was also considered
      by S.~Dimopoulos and J.~March-Russell~\cite{private}.}
is that the coupling of the right-handed neutrino $N$ to the ordinary
neutrino $\nu_L$ is automatically suppressed by a {\it volume factor}\/
corresponding to the extra compactified dimension.
  Such a volume factor arises from the normalization of the wavefunction
  of the $N$ field in the compactified dimension, and will be discussed
further below.
This volume factor can also provide a natural mechanism
for suppressing the Yukawa coupling and yielding a light neutrino
mass.

Both of the above mechanisms suppress the neutrino mass
by directly suppressing the value of $m$.
However, as we shall now discuss, it may also be possible
to suppress the neutrino mass via a higher-dimensional analogue of the
seesaw mechanism.

Once again, we shall assume that the right-handed neutrino feels
extra dimensions, while the left-handed neutrino $\nu_L$ does not.
Specifically, in higher dimensions (\eg, in five dimensions, for
concreteness), we consider
a Dirac fermion $\Psi$, which in the Weyl basis can be decomposed
into two two-component spinors:  $\Psi = (\psi_1,\bar\psi_2)^T$.
When the extra spacetime dimension is compactified on
a $\IZ_2$ orbifold, it is natural for one of the two-component
Weyl spinors, \eg, $\psi_1$, to be taken to be even
under the $\IZ_2$ action $y\to -y$, while the other spinor
$\psi_2$ is taken to be odd.
If the left-handed neutrino $\nu_L$ is restricted
to a brane located at the orbifold fixed point $y=0$,
then $\psi_2$ vanishes at this point and so the most
natural coupling is between $\nu_L$ and $\psi_1$.
For generality, we will also include
a possible ``bare'' Majorana mass $M_0$ for $\Psi$.
This then results in a Lagrangian of the form
\beqn
  {\cal L} &=& \int d^{4} x \,dy ~M_s\, \biggl\lbrace
     {\bar\psi}_1 i{\bar\sigma}^\mu \partial_\mu \psi_1
       +{\bar\psi}_2 i{\bar\sigma}^\mu \partial_\mu \psi_2
      + \half M_0 \left(
      \psi_1 \psi_1 + \psi_2 \psi_2 + {\rm h.c.} \right)\biggr\rbrace
                                 \nonumber\\
      && +~\int d^4 x ~ \biggl\lbrace
         {\bar\nu}_L i{\bar\sigma}^\mu D_\mu \nu_L
         ~+~({\hat m} \nu_L \psi_1|_{y=0} + {\rm h.c.})     \biggr\rbrace~.
\label{klag}
\eeqn
Here $y$ is the coordinate
of the extra compactified spacetime dimension,
and $M_s$ is the
mass scale of the higher-dimensional fundamental
theory (\eg, a reduced Type~I string scale).
The first line represents the kinetic-energy term for
the five-dimensional $\Psi$ field
as well as the bare Majorana mass term
$\half M_0 \bar\Psi^{\rm c}\Psi$.
By contrast, the second line represents the kinetic energy
of the four-dimensional two-component neutrino field $\nu_L$
as well as the coupling between $\nu_L$ and $\psi_1$.
Note that in five dimensions, a bare Dirac mass term for $\Psi$
would not have been invariant under the action of the $\IZ_2$ orbifold,
since $ {\bar \Psi}\Psi\sim \psi_1\psi_2 +$ h.c.

Next, we compactify the Lagrangian (\ref{klag}) down to four dimensions
by expanding the five-dimensional $\Psi$ field in Kaluza-Klein
modes.  Imposing the orbifold relations
$\psi_{1,2}(-y)=\pm \psi_{1,2}(y)$
implies that our Kaluza-Klein decomposition takes the form
\beqn
     \psi_1(x,y) &=& {1\over \sqrt{2\pi R}}\,\sum_{n=0}^\infty
         \psi_1^{(n)}(x)\,\cos (ny/R)\nonumber\\
     \psi_2(x,y) &=& {1\over \sqrt{2\pi R}}\,\sum_{n=1}^\infty
         \psi_2^{(n)}(x)\,\sin (ny/R)~.
\label{KKdecomp}
\eeqn
For convenience, we shall also define the linear combinations
$N^{(n)}\equiv(\psi_1^{(n)}+\psi_2^{(n)})/\sqrt{2}$
and
$M^{(n)}\equiv(\psi_1^{(n)}-\psi_2^{(n)})/\sqrt{2}$
for all $n>0$.
Inserting (\ref{KKdecomp}) into (\ref{klag}) and
integrating over the compactified dimension
then yields
\beqn
  {\cal L} &=&\int d^4 x
     ~ \Biggl\lbrace
    {\bar\nu}_L i{\bar\sigma}^\mu D_\mu \nu_L
     + {\bar \psi}_1^{(0)} i{\bar\sigma}^\mu\partial_\mu \psi_1^{(0)}
     + \sum_{n=1}^\infty \left(
     {\bar N}^{(n)} i{\bar\sigma}^\mu\partial_\mu N^{(n)}
     +{\bar M}^{(n)} i{\bar\sigma}^\mu\partial_\mu M^{(n)} \right) \nonumber\\
    && ~~~~+~ \biggl\lbrace \half \, M_0 \,\psi_1^{(0)} \psi_1^{(0)}
        ~+~
         \half \sum_{n=1}^\infty \, \left\lbrack
         \left(M_0 + {n\over R}\right) N^{(n)} N^{(n)}
         + \left(M_0 - {n\over R}\right) M^{(n)} M^{(n)} \right\rbrack
            \nonumber\\
    && ~~~~+~ m \nu_L \psi_1^{(0)} +
           \sum_{n=1}^\infty
         \left( m_N^{(n)}\, \nu_L N^{(n)}
         +  m_M^{(n)}\, \nu_L  M^{(n)}\right) ~+~ {\rm
h.c.}\biggr\rbrace\Biggr\rbrace~.
\label{tglag}
\eeqn
Here the first line gives the four-dimensional kinetic-energy terms,
while the second line
gives the Kaluza-Klein and Majorana mass terms.
Note that the Kaluza-Klein masses $n/R$ are replaced by
$(n_1+in_2)/R$ in the case of {\it two} extra spacetime dimensions.\footnote{
     At first glance, it may seem surprising that a {\it complex}\/
     Kaluza-Klein mass is generated for $\delta \geq 2$.  However,
     only the modulus $\sqrt{n_1^2 + n_2^2}/R$ is the physical mass.
     The complex phase arises because the fermionic Kaluza-Klein
     reduction can yield only mass terms which are linear in the Kaluza-Klein
     momenta $n_i/R$. }
The third line of (\ref{tglag}) describes the coupling
between the four-dimensional neutrino $\nu_L$
and the five-dimensional field $\Psi$.
Note that in obtaining this Lagrangian, it is necessary to rescale
the individual $\psi_1^{(0)}$, $N^{(n)}$, and $M^{(n)}$ Kaluza-Klein modes
so that their four-dimensional kinetic-energy terms are canonically normalized.
This then results in a suppression of the
Dirac neutrino mass coupling $\hat m$
by the factor $(2\pi M_s R)^{1/2}$.
In the third line, we have therefore simply defined the effective Dirac
neutrino mass
couplings $m_N^{(n)}=m_M^{(n)}=m$  for all $n$, where
\beq
            m ~\equiv~  {{\hat m}\over \sqrt{2}\,\sqrt{\pi M_sR}}~.
\label{mdef}
\eeq
For a general $\IZ_N$ orbifold and $\delta$ extra dimensions,
this volume factor $(\pi R)^{1/2}$ generalizes to $(2\pi R/N)^{\delta/2}$.

One important dimensionless number in our analysis will be
the product $mR$.
Therefore, let us give a rough estimate.
First, we note that for a $\IZ_2$ orbifold, we have
\beq
   mR ~\sim~ {1\over (2\pi)^{\delta/2}}\left({ \hat m \over  M_s}\right)\,
     {1\over (M_s R)^{\delta/2-1}} ~.
\label{mRvalue}
\eeq
Regardless of whether the extra dimensions are
``universal'' or are felt only by gravity,
we always have $M_s R >1$;  indeed, in the latter case we even
have $M_s R\gg 1$.
Likewise we expect $\hat m / M_s <1$.
Thus, for $\delta\geq 2$, we find that
$mR \ll 1$ in all cases, with this approximation
becoming particularly appropriate in the case of
gravity-only extra dimensions.
For example, in the case of gravity-only extra dimensions,
we have $mR \leq {\cal O}(10^{-3})$ for $\delta=2$.
Note that the $\delta=1$ case is also of some interest.
Although the case of one gravity-only extra dimension
is excluded experimentally if the Planck scale is pushed to the
TeV-range, it nevertheless remains possible for the higher-dimensional
$\Psi$ field to feel only {\it some}\/ of the large extra dimensions.  This
would depend on the sector from which the $\Psi$ field originates
in the Type~I string theory.  In such cases, $mR$ can be quite a bit larger.
For example, with $n$ total extra dimensions and $\delta=1$ (\ie, only
one of these extra dimensions felt by the $\Psi$ field), we have
$mR \geq {\cal O}(1)$, with $mR$ ranging from
${\cal O}(10^5)$ for $n=2$ to ${\cal O}(1)$ for $n=6$.
It is also possible (and indeed suggested~\cite{DDG,DDGR,Antoniadis}) that
the total compactification space may be anisotropic, with some
extra dimensions large and others small.  This would then alter
the above estimates significantly.
Thus, although we shall often focus on the case $mR\ll 1$,
we shall attempt to keep our discussion general.

Given the Lagrangian (\ref{tglag}), we see that
the Standard-Model neutrino
$\nu_L$ will mix with the entire tower of Kaluza-Klein states
of the higher-dimensional $\Psi$ field.
Indeed, if we restrict our attention for the moment to the case of only one
extra dimension for simplicity and define
\beq
        \calN^T ~\equiv~ (\nu_L, \psi_1^{(0)},
                  N^{(1)}, M^{(1)},
                  N^{(2)}, M^{(2)}, ...)~,
\label{calNdef}
\eeq
we see that the mass terms in the Lagrangian (\ref{tglag})
take the form $\half(\calN^T \calM \calN+{\rm h.c.})$
where the mass matrix is symmetric and takes the form
\beq
      \calM ~=~ \pmatrix{
         0 &  m   &   m^{(1)}_N  &   m^{(1)}_M  &   m^{(2)}_N  &  m^{(2)}_M  &
\ldots \cr
         m &  M_0 &   0  &   0  &   0  &  0  & \ldots \cr
       m^{(1)}_N &  0   &   M_0+1/R  &   0  &   0  &  0  & \ldots \cr
       m^{(1)}_M &  0   &   0  &   M_0-1/R  &   0  &  0  & \ldots \cr
       m^{(2)}_N &  0   &   0  &   0  &   M_0+2/R  &  0  & \ldots \cr
       m^{(2)}_M &  0   &   0  &   0  &   0   &  M_0-2/R  & \ldots \cr
         \vdots  &  \vdots &   \vdots  &   \vdots &   \vdots &  \vdots  &
\ddots \cr}~.
\label{newmatrix}
\eeq

Before proceeding further,
let us discuss the assumptions inherent in the form of this
mass matrix.
First, note that
the entries $m,m_N^{(n)},m_M^{(n)}$
reflect the coupling in (\ref{tglag})
between the left-handed neutrino state
(which feels only four spacetime dimensions)
and the $\Psi$ field (which also feels the extra dimensions).
As we have seen above, the condition $m=m_N^{(n)}=m_M^{(n)}$
results for the case of a straightforward coupling of $\nu_L$
(restricted to a brane located at $y=0$) to the
higher-dimensional $\psi_1$ field.
However, as we shall discuss in Sect.~2.5,
it is possible to consider more general brane configurations
in which the parameters $m$, $m^{(n)}_N$, and $m^{(n)}_M$ are all unequal.
We shall therefore leave these couplings completely general,
as in (\ref{newmatrix}).
Of course, the {\it value}\/ of these couplings is no longer to
be associated
with the value given in (\ref{Higgsvev}), since the presence of the
extra dimensions alters the result given
in (\ref{Higgsvev}) by an overall volume factor, as discussed above.
Second, note that the remaining entries along the diagonal
reflect the
masses of the Kaluza-Klein modes of the $\Psi$ field, as given in
(\ref{KKmasses}).  The contribution from the bare Majorana mass
$M_0$ is also included.
Third, note that we have not introduced any additional off-diagonal
non-zero entries in this mass matrix, for
such non-zero entries would violate
Kaluza-Klein momentum conservation.
It might seem at first that conservation of Kaluza-Klein momentum
would also forbid the couplings
between the left-handed neutrino $\nu_L$ and the excited Kaluza-Klein
modes of the higher-dimensional $\Psi$ field.
However,
the difference in this case is the fact that the left-handed
neutrino is presumed not to feel the extra spacetime dimensions,
and is therefore essentially restricted to a brane with respect
to these extra dimensions.
Kaluza-Klein momentum conservation
therefore does not apply
for such couplings
because the presence of the brane breaks translational invariance in the
compactified direction(s).
Thus, we conclude that the most general form
for the mass matrix is the one given
in (\ref{newmatrix}).

We shall now proceed to study the physical implications of
this mass matrix.  Most of our attention will focus on the generation
of a {\it seesaw mechanism}\/, just as in the usual four-dimensional case.
This is important for the following reason.
Let us imagine, for the moment, that the bare Majorana mass is absent,
so that $M_0=0$.   By itself, the usual four-dimensional
seesaw mechanism between $\nu_L$ and the zero-mode field $\psi_1^{(0)}$
would then result in a $2\times 2$ matrix
\beq
       \pmatrix{ 0 & m \cr m &  0}~,
\label{noseesaw}
\eeq
which leads to the degenerate eigenvalues $\lambda_\pm = \pm m$.
These eigenvalues can be combined to form a Dirac mass for the neutrino.
Comparing with (\ref{usualseesaw}), we see that this
is the limit in which there is no seesaw at all --- \ie, the
two lightest states $\nu_L$ and $\psi^{(0)}_1$ remain degenerate,
with neither becoming lighter or heavier than the other.
However, as we shall see below, this need not
remain the case in higher dimensions:
the excited Kaluza-Klein states can induce a seesaw even
if ground-state zero-mode itself does not.
This will be important if we want to give the neutrino
a {\it Majorana}\/ mass, rather than merely a Dirac mass.
As we shall see, the degree to which a true seesaw mechanism can
be realized ultimately depends on the values of
\beq
       \left[m_N^{(n)}\right]^2-\left[m_M^{(n)}\right]^2~.
\eeq
We shall therefore begin by studying the physical implications
of our mixing matrix (\ref{newmatrix}) in a number of simple limits
in order to elucidate its basic properties and consequences for
the resulting neutrino mass.

\subsection{A toy model:  ~The case
          $M_0 = m_M^{(n)} = 0$, ~$m_N^{(n)}=m$}

In order to most dramatically illustrate the possibilities of
a higher-dimensional seesaw mechanism, let us begin by
considering the extreme case of (\ref{newmatrix})
in which $m_M^{(n)}=0$ for all $n$.
As we shall see, this will result in the strongest seesaw behavior ---
\ie, the maximal splitting between the light eigenvalues $\lambda_\pm$.
Under the assumption that
$m_M^{(n)}=0$ for all $n$,
the neutrino $\nu_L$ no longer couples
to the $M^{(n)}$ modes, and hence they decouple from the problem.
For simplicity, we shall also
disregard the bare Majorana mass, setting $M_0=0$.
This then results in the simplified Lagrangian
\beqn
  {\cal L} &=&\int d^4 x
     ~ \Biggl\lbrace
    {\bar\nu}_L i{\bar\sigma}^\mu D_\mu \nu_L
   ~+~ \sum_{n=0}^\infty {\bar N}^{(n)} i{\bar\sigma}^\mu\partial_\mu
   N^{(n)} \nonumber\\
  && ~~+~ \biggl\lbrace
       \half\, \sum_{n=1}^\infty \,{n\over R} \,{N}^{(n)}N^{(n)}
      ~+~ \sum_{n=0}^\infty \,m\, {\nu}_L N^{(n)} +{\rm h.c.} \biggr\rbrace
           \Biggr\rbrace ~
\label{tglagold}
\end{eqnarray}
where we have defined $N^{(0)}\equiv \psi_1^{(0)}$.
We emphasize that in writing (\ref{tglagold}),
we have neglected the terms
containing the (decoupled) field $M^{(n)}$.
Although it may seem that lepton number is apparently broken
in the Kaluza-Klein mass terms for
$N^{(n)}$, this is precisely compensated for by the analogous mass
terms for $M^{(n)}$ as in (\ref{tglag}).
Thus, defining the reduced set of fields
\beq
        \calN^T ~\equiv~ (\nu_L, N^{(0)}, N^{(1)}, N^{(2)}, ...)~,
\label{calNdefnew}
\eeq
we see that the mass terms in the Lagrangian (\ref{tglagold})
correspond to the simplified mass matrix
\beq
      \calM ~=~ \pmatrix{
         0 &  m   &   m  &   m  &   m  &  m  & \ldots \cr
         m &  0   &   0  &   0  &   0  &  0  & \ldots \cr
         m &  0   &   1/R  &   0  &   0  &  0  & \ldots \cr
         m &  0   &   0  &   2/R  &   0  &  0  & \ldots \cr
         m &  0   &   0  &   0  &   3/R  &  0  & \ldots \cr
         m &  0   &   0  &   0  &   0   &  4/R  & \ldots \cr
         \vdots  &  \vdots &   \vdots  &   \vdots &   \vdots &  \vdots  &
\ddots
 \cr}~.
\label{newmatrixreally}
\eeq

The remarkable feature of (\ref{newmatrixreally})
is that the resulting ``light'' eigenvalues $\lambda_\pm$
are {\it maximally split from each other}\/, with a splitting
that becomes infinitely great as the size of the matrix is
taken to infinity.
In particular, while one eigenvalue grows infinitely heavy
as more and more of the Kaluza-Klein states participate
in the mixing, the other becomes arbitrarily light.
We will prove this statement analytically below, but
let us first sketch how this happens in practice.
Once again, we begin by focusing on only the
upper-left $2\times 2$ mixing sub-matrix
between $\nu_L$ and $N^{(0)}$ alone.
This is the same as the matrix (\ref{noseesaw}),
which produces two eigenvalues $\pm m$.
Therefore both of the resulting mass eigenstates
would have masses equal to $m$  --- \ie, there would be no seesaw
between $\nu_L$ and $N^{(0)}$.
However, as we increase the size of this matrix by adding the further
rows and columns corresponding to the excited Kaluza-Klein states,
we find that the cumulative effect of the excited Kaluza-Klein states is to
pull the negative eigenvalue $-m$ further in the negative direction,
but also to decrease the positive eigenvalue $+m$.
Ultimately, as the dimensionality of this mass matrix is taken
to infinity, the negative eigenvalue $-m$ falls all the way to negative
infinity while the positive eigenvalue $+m$ falls all the way to zero.
Note that each new row/column also introduces a new eigenvalue
which, in the limit as the matrix becomes infinite-dimensional,
simply remains fixed near $n/R$.
Thus, we find that our infinite-dimensional matrix produces
one zero eigenvalue and one infinite eigenvalue,
with all other eigenvalues of size $R^{-1}$ or larger.

We shall now give an analytical derivation of the eigenvalues of this
infinite-dimensional matrix.
We begin by considering a matrix of finite size $(n+2)\times (n+2)$,
so that the highest diagonal entry is $n/R$.  This matrix will therefore
have $(n+2)$ different
eigenvalues;  these consist of the $n$ different eigenvalues
$\lambda_k$ ($k=1,...,n$)  corresponding to the
excited Kaluza-Klein states, as well as the two remaining ``light''
eigenvalues $\lambda_+$ and $\lambda_-$ whose values are
respectively $\pm m$ in the special case $n=0$, as discussed above.
Our procedure will be to solve for the
``light'' eigenvalues $\lambda_\pm$ as a function of $n$,
and to consider their behavior as $n\to \infty$.

We begin by considering the characteristic eigenvalue equation
${\rm det}(\calM -\lambda I)=0$.  Given the mass matrix $\calM$ in
(\ref{newmatrixreally}), this equation takes the exact
analytic form
\beq
   \left\lbrack \prod_{k=1}^n \left({k\over R}-\lambda\right)
         \right\rbrack \,
    \left[\lambda^2-m^2 + \lambda m^2 R\sum_{k=1}^n
    {1\over k-\lambda R }\right] ~=~ 0~.
\label{tgcheq}
\eeq
However, since we know that we can always ultimately
write this eigenvalue equation in the form
\beq
          \left\lbrack \prod_{k=1}^n (\lambda_k-\lambda) \right\rbrack
                  \, \left(\lambda_+-\lambda\right)
                  \, \left(\lambda_--\lambda\right)     ~=~0~,
\eeq
we see that we can obtain a number of different relations
amongst the eigenvalues by considering the coefficients
of various powers of $\lambda$ in (\ref{tgcheq}).  For example,
the constant term $C_0$ (\ie, the coefficient of $\lambda^0$)
gives the product of the eigenvalues, $\prod \lambda$,
which is nothing but
the determinant of ${\cal M}$.  Likewise, the
coefficient $C_1$
of the term linear in $\lambda$ is identified as
the sum of the products of all possible subsets of
$n+1$ of the eigenvalues, \ie,
$C_1= -\sum_{i_1\dots i_{n+1}} \lambda_{i_1}\dots\lambda_{i_{n+1}}$,
where the $i$-indices run over the set $\lbrace 1,2,...,n,+,-\rbrace$.
Similarly, the coefficient $C_{n+1}$ of the $\lambda^{n+1}$
term gives $(-1)^{n+1}\sum \lambda$, which is equivalently
$(-1)^{n+1}$ times the trace of the matrix $\calM$.

By examining the matrix ${\cal M}$ and the characteristic
equation (\ref{tgcheq}), it is easy to see that
\beq
     C_0 ~=~ - {n! \,m^2\over R^n}~,~~~~
     C_1 ~=~ {2 \,n!\, m^2 \over R^{n-1}} \, \sum_{k=1}^n {1\over k}~,~~~~
     C_{n+1} ~=~ (-1)^{n+1} {1\over R}\, \sum_{k=1}^n \,k~.
\eeq
Moreover, for $mR\ll 1$, it is easy to show that the excited
Kaluza-Klein eigenvalues behave as $\lambda_k \approx k/R + m^2R/k +...$
for large $n$.  (This will be discussed further below.)
Using this information, we can then obtain various simultaneous equations
for $\lambda_+$ and $\lambda_-$.
For example, from the $C_0$ determinant relation we find
\beq
      \lambda_+\lambda_- ~\approx~ -m^2 \left(
      1+m^2R^2\sum_{k=1}^n {1\over k^2} +\dots \right)^{-1}~.
\label{tgeq2}
\eeq
Likewise, from the $C_1$ relation we find
\beq
     \lambda_+ ~+~ \lambda_- ~+~ R\,\lambda_+\lambda_-\,\sum_{k=1}^n
           {1\over k}
     ~\approx~ -2 m^2 R\sum_{k=1}^n {1\over k}~,
\label{tgeq1}
\eeq
and from the $C_{n+1}$ trace relation we find
\beq
    \lambda_+ ~+~ \lambda_- ~\approx~ -m^2 R\,\sum_{k=1}^n {1\over k}~.
\label{tgeq3}
\eeq

We can now solve any two of these equations
simultaneously for the eigenvalues $\lambda_\pm$.
In all cases, we obtain
\beq
     \lambda_\pm ~=~ \half  \left[\mu \pm \sqrt{ \mu^2 +4m^2} \,\right]~
\label{tgeigpoly}
\eeq
where we have defined the quantity
\beq
           \mu ~\equiv~ -m^2 R \sum_{k=1}^n {1\over k} ~.
\label{mudef}
\eeq
Thus, comparing with the usual seesaw result in (\ref{usualseesaw}),
we see that the entire tower of Kaluza-Klein states has generated
a seesaw that splits the two lightest eigenvalues.  We shall discuss the
physical interpretation of $\mu$ below.

There are two limits that will be of interest to us, depending
on the size of $\mu$.
The first case, with $mR\ll 1$ and finite $n$, corresponds to
$\mu \ll m$.  In this case, we obtain the solutions
\beq
      \lambda_\pm ~=~ \pm m ~-~\half \,m^2 R \, \sum_{k=1}^n {1\over k}~.
\label{finitenresults}
\eeq
As discussed
above, these are the two light eigenvalues that arise when only the
lightest Kaluza-Klein states participate in the mixing.
By contrast, the full ``seesaw'' limit arises as we take
$n\to \infty$, with all Kaluza-Klein states included in the
mixing.  In this case, we have $\mu \gg m$, whereupon we find
\beq
      \lambda_+ ~\sim~  {1\over R \sum_{k=1}^n 1/k}
      ~\sim~  {1\over R \ln n}~
\label{tgzero}
\eeq
and
\beq
     \lambda_- ~\sim~ -\,m^2 R\, \sum_{k=1}^n {1\over k}
            ~\sim~ -\,m^2 R\, \ln n~ .
\label{tgzerotwo}
\eeq
We thus see that $\lambda_+$ becomes arbitrarily light
as $n\to\infty$, while $\lambda_-$ becomes arbitrarily heavy.
It is interesting that the $n\to\infty$ limit is capable of
producing such an infinite splitting between these two eigenvalues ---
and with it an arbitrarily light eigenvalue  ---
regardless of the intrinsic value of the radius $R$.

Note that these results
can also be seen
directly from the characteristic equation (\ref{tgcheq}),
 {\it regardless of the value of $m R$}\/,
by noticing that $\lambda\to 0$
is a solution of the term in square brackets.
Specifically, near $\lambda=0$, the sum
$\sum_{k=1}^n 1/(k -\lambda R)$ is analytic in $\lambda$. Thus,
performing a Taylor
expansion about the origin and keeping only
terms linear in $\lambda$,
we again find a solution that behaves
like (\ref{tgzero}).
This then implies the solution (\ref{tgzerotwo}).
This argument does not rely on the value of $m R$,
and thus we see that these solutions
continue to hold regardless of the value
of $mR$.
Similarly, note that the excited Kaluza-Klein eigenvalues given above,
namely $\lambda_k \approx k/R + m^2 R/k$, are valid only for
finite $n$.  In the limit $n\rightarrow\infty$, one finds that
$\lambda_k \sim k/R +1/(R\ln n)$. This can be seen by
substituting the value $\lambda_k = k/R + c$ into the
characteristic equation (\ref{tgcheq}) and showing, in a
fashion similar to that for the zero eigenvalue, that the
term in square brackets vanishes if $c\sim 1/(R\ln n)$.

\begin{figure}[ht]
\centerline{ \epsfxsize 3.5 truein \epsfbox {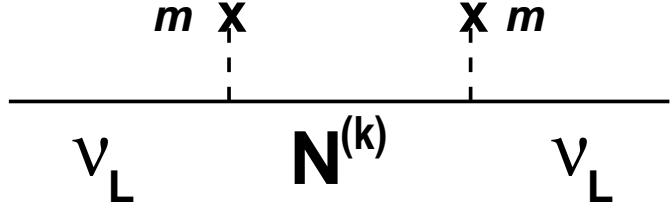}}
\caption{Tree-level diagram showing the generation of an effective
          neutrino mass term through a mixing with the right-handed
          Kaluza-Klein states $N^{(k)}$.  }
\label{newfigure}
\end{figure}

Another useful way to derive these results
is to consider the tree-level Feynman diagram in Fig.~\ref{newfigure}.
For any given Kaluza-Klein state $N^{(k)}$, this diagram can be
interpreted as contributing to an individual seesaw between $\nu_L$ and
$N^{(k)}$.
In the limit $m\ll 1/R$, the masses of the excited Kaluza-Klein states
exceed the size of their Dirac couplings $m$, and can be integrated out.
Thus, by summing over all possible intermediate Kaluza-Klein states (\ie,
by summing over all of the individual Kaluza-Klein seesaw contributions),
we then generate a neutrino mass term $\mu \nu_L \nu_L$, where
\beq
       \mu ~=~ \sum_{k=1}^n m{-1\over k/R} m ~\approx~ -m^2  R \ln n~.
\eeq
This is the same quantity defined in (\ref{mudef}).
Once these massive Kaluza-Klein states are integrated out,
the problem is reduced to an effective seesaw
mechanism between $\nu_L$ and $N^{(0)}$, with mass matrix
\beq
            \pmatrix { \mu & m \cr
                       m & 0 \cr}~.
\label{twobytwo}
\eeq
This then leads to the eigenvalues given in (\ref{tgeigpoly}).
We also note, in passing,
that Fig.~\ref{newfigure} also gives a contribution to the wavefunction
renormalization of the left-handed neutrino.
The corresponding Kaluza-Klein summation is finite for
$\delta=1$, logarithmically divergent for $\delta=2$, and power-law divergent
for $\delta >2$.  As we shall discuss, however, the Kaluza-Klein sum must be
truncated at the string scale, and therefore the final result is a finite
renormalization factor which we shall implicitly disregard.

One important point that emerges from the above discussion is that
this higher-dimensional seesaw mechanism is {\it inverted}\/ relative
to the usual four-dimensional one.  This is clear, for example, upon
comparing (\ref{twobytwo}) with (\ref{usualmatrix}).
Specifically, it is the neutrino $\nu_L$
that becomes heavier as  more and more
Kaluza-Klein states are included in the mixing,
while the zero-mode $N^{(0)}$ becomes light.
In other words, while $\mu$ serves as an effective seesaw scale for the light
eigenstate $N^{(0)}$, the effective seesaw scale for the
neutrino eigenstate is actually
\beq
       M_{\rm eff} ~=~ - {m^2 \over \mu} ~=~  {1\over R \ln n}~.
\label{Meff}
\eeq
Thus, our seesaw mechanism rotates the states in a direction that
is opposite to the direction that usually emerges in the four-dimensional
case.

However, this may be useful for the following reason.
Recall that the approach we have followed is one based on an effective
field theory.  If our true underlying theory is a string theory
with mass scale $M_s$, then
we expect our considerations to be valid only up to
the mass scale $M_s$.
This in turn means that the maximum number of Kaluza-Klein
states which should enter into our considerations is
$n_{\rm max}\sim {\cal O}(R M_s)$.
Given this observation, it is natural to identify
the result (\ref{tgzerotwo})
with values of the neutrino masses suggested
by the recent SuperKamiokande data.
However, as we discussed in Sect.~2.2,
there is tremendous variation in the size of $mR$ and in the
resulting rescaled coupling $m$ defined in (\ref{mdef}).
Indeed, it generically appears to be difficult to arrange $m$ itself
to be of the size suggested by the SuperKamiokande results.
Thus, we now see that our seesaw mechanism plays an important role
by offering the possibility of {\it altering}\/ this mass:
\beq
         m ~\longrightarrow~ m^2 R \ln ( R M_s)~.
\eeq
In fact, in many instances $m$ itself is too {\it small}\/ to
agree with these neutrino masses.  The above substitution
may therefore enable a useful {\it enhancement}\/ of the na\"\i ve
value of $m$.
Of course, the phenomenological details of this
mechanism ultimately depend
on the sizes of $mR$ and $RM_s$, which in turn depend
crucially on the geometry (and in particular the anisotropy)
of the compactification manifold.

We also stress that in this section we have been considering
only the illustrative toy model that emerges from the extreme limit in which
we take $m_M^{(n)}=0$ for all $n$.  Nevertheless, many of the crucial
features of this scenario will continue to hold in the subsequent
scenarios that have a more natural realization in string theory.

\subsection{The orbifold case:  ~$m=m_N^{(n)}=m_M^{(n)}$, $M_0=0$}

Let us now return to (\ref{newmatrix}), and discuss the
special case in which the Dirac couplings satisfy
$m=m_N^{(n)}=m_M^{(n)}$ for all $n$.
As discussed in Sect.~2.2, this corresponds to a straightforward
orbifold coupling between the four-dimensional $\nu_L$ field
and the higher-dimensional $\psi_1$ field.
For simplicity, we shall again disregard the possible bare Majorana
mass term, setting $M_0=0$.
In this situation,
our mass matrix (\ref{newmatrix}) then becomes
\beq
      \calM ~=~ \pmatrix{
         0 &  m   &   m  &   m  &   m  &  m  & \ldots \cr
         m &  0   &   0  &   0  &   0  &  0  & \ldots \cr
         m &  0   &   1/R  &   0  &   0  &  0  & \ldots \cr
         m &  0   &   0  &   -1/R  &   0  &  0  & \ldots \cr
         m &  0   &   0  &   0  &   2/R  &  0  & \ldots \cr
         m &  0   &   0  &   0  &   0   &  -2/R  & \ldots \cr
         \vdots  &  \vdots &   \vdots  &   \vdots &   \vdots &
         \vdots  & \ddots \cr}~.
\label{tgnewmatrix}
\eeq
The characteristic polynomial equation which determines
the eigenvalues of this mass matrix then takes the form
\beq
   \left\lbrack \prod_{k=1}^\infty \left({k^2\over R^2}-\lambda^2\right)
         \right\rbrack \,
    \left[\lambda^2-m^2 +2 \lambda^2 m^2 R^2\sum_{k=1}^\infty
    {1\over k^2-\lambda^2 R^2 }\right] ~=~ 0~,
\label{tgnewcheq}
\eeq
which is invariant under $\lambda\to -\lambda$.  From this
we immediately see that all eigenvalues are exactly {\it degenerate},
falling into pairs of opposite sign.
This implies, in particular, that the two lightest eigenvalues are
degenerate (combining to produce a Dirac mass for the neutrino),
and that there is no seesaw behavior.
Indeed, all of the resulting masses for the Kaluza-Klein eigenstates
are Dirac as well.
This is ultimately a consequence of the alternating signs in the
diagonal entries of (\ref{tgnewmatrix}).

\begin{figure}[t]
\centerline{
      \epsfxsize 3.1 truein \epsfbox {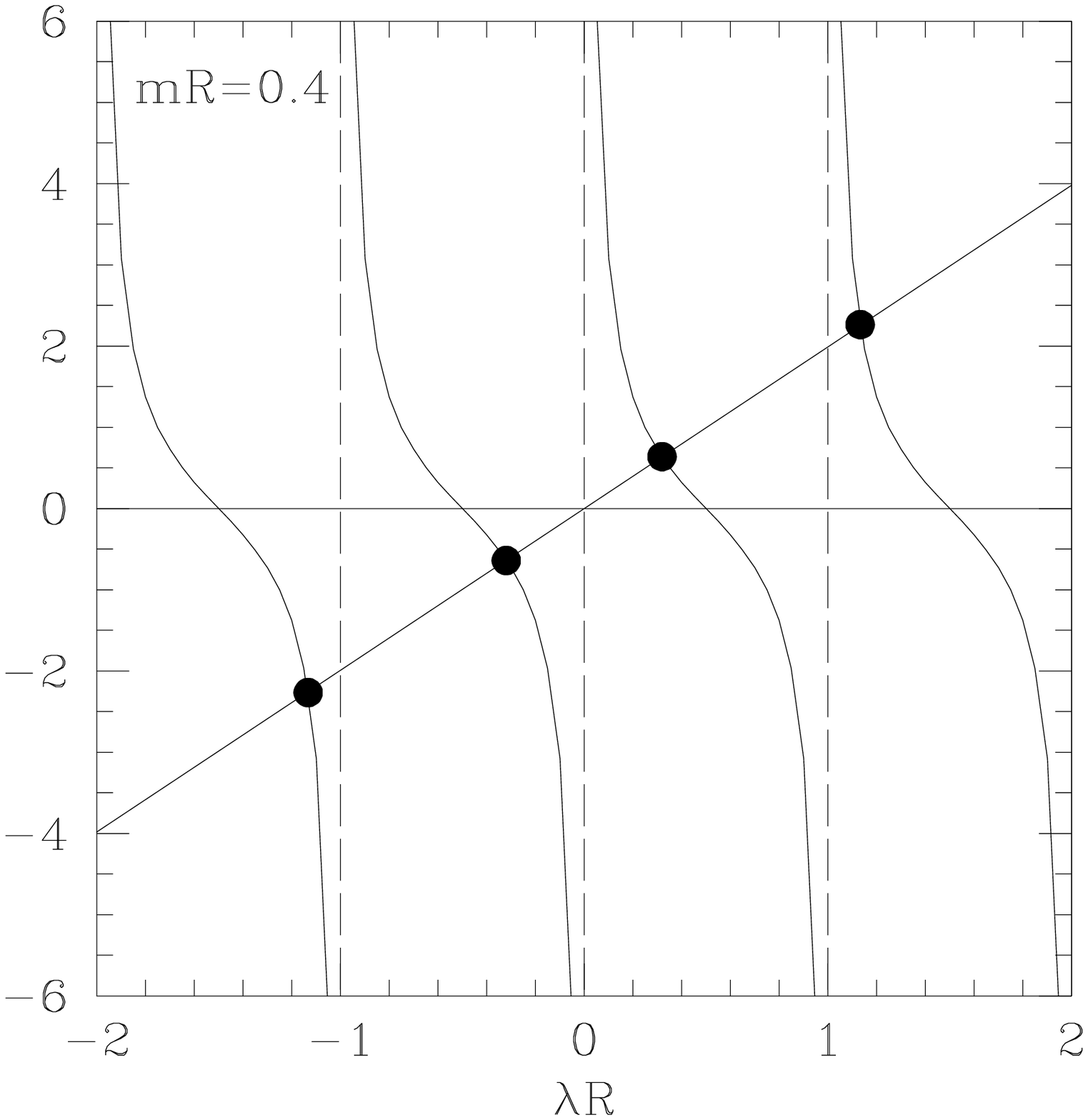}
      \hskip 0.2 truein
      \epsfxsize 3.1 truein \epsfbox {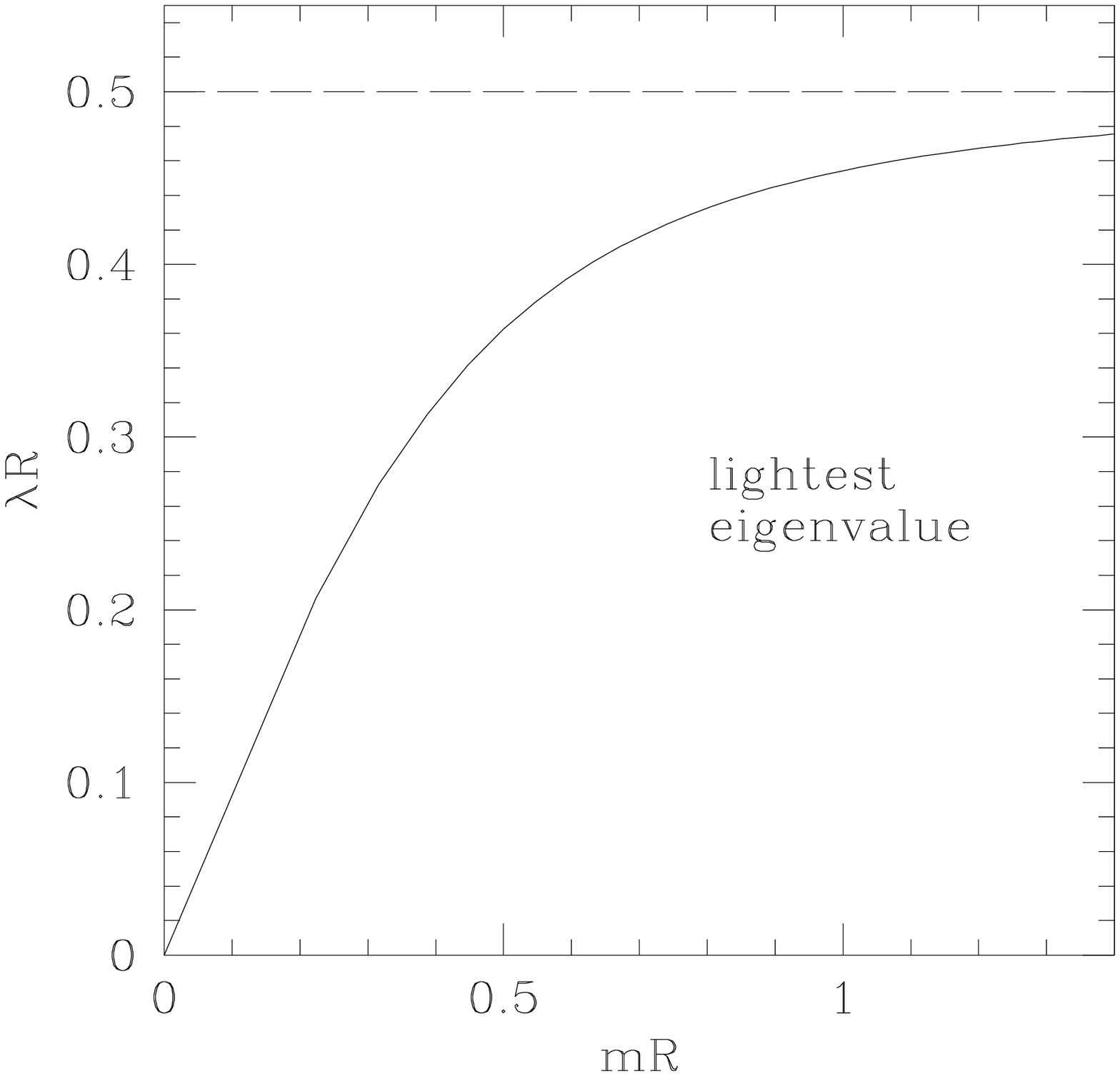}}
\caption{  (a)  Eigenvalue solutions to (\protect\ref{trans1}),
           represented as those values of $\lambda$ for which
          $\cot(\pi \lambda R)$ intersects
         $\lambda R/ [\pi (mR)^2]$.  We have taken the
               fixed value $mR=0.4$ for this plot.
           The behavior of the eigenvalues as functions of $mR$
          can be determined graphically by changing the slope
           of the intersecting diagonal line.
          (b)  The lightest eigenvalue (neutrino mass) $\lambda_+$
                as a function of $mR$.  For $mR\ll 1$, we see that
              the curve is approximately linear,
            corresponding to $\lambda_+ \approx m$.  However,
            as $mR$ increases, the neutrino mass increases non-linearly,
         ultimately reaching an asymptote at $\lambda_+ R =1/2$ at which
         point the volume factor becomes irrelevant.}
\label{orbfigures}
\end{figure}

In order to solve this eigenvalue equation, it is convenient to note
that $\lambda = k/R$ is never a solution (unless of course $m=0$),
as the cancellation that would occur in the first factor
in (\ref{tgnewcheq})
is offset by the divergence of the second factor.
We are therefore free to disregard the first factor entirely, and
focus on solutions for which the second factor vanishes.
The summation in second factor can then be performed exactly, resulting
in the transcendental equation
\beq
          \lambda R ~=~ \pi (m R)^2 \,\cot ( \pi \lambda R)~.
\label{trans1}
\eeq
All of the eigenvalues can be determined from this equation, as functions
of the product $mR$.  The solutions are shown graphically in
Fig.~\ref{orbfigures}(a).
We immediately see that in the limit $mR\to 0$ (corresponding to $m\to 0$),
the eigenvalues are $k/R$, $k\in \IZ$, with a double eigenvalue at $k=0$.
Conversely, in the limit $mR\to\infty$, the eigenvalues with $k>0$ smoothly
shift to $(k+\half)/R$, while those with $k<0$ shift to $(k-\half)/R$ and
the double zero eigenvalue
splits towards the values $\pm 1/(2R)$.
In order to derive general analytical expressions valid
in the limit $mR\ll 1$, we can solve (\ref{trans1}) iteratively by
power-expanding the cotangent function.
To order ${\cal O}(m^5 R^5)$, this gives the solutions
\beq
         \lambda_{\pm k} ~=~ \pm {k\over R} ~+~
       {1\over 2 R(1+ \pi^2 m^2 R^2/ 3)} \left\lbrack
            \mp k \pm \sqrt{ k^2 + 4 m^2 R^2 (1+ \pi^2 m^2 R^2
/3)}\right\rbrack~,~~~~~
            k\in \IZ~,
\label{gensoln}
\eeq
where $\lambda_{\pm k}$ are the two eigenvalues at each Kaluza-Klein level $k$.
Note that this expression also includes the ``light'' eigenvalues $\lambda_\pm$
at $k=0$, as shown in Fig.~\ref{orbfigures}(b).
Expanding to order ${\cal O}(m^5 R^5)$, we thus find
\beq
  \lambda_{\pm} ~=~ \pm m\,\left( 1- {\pi^2 \over 6} m^2 R^2 + ...\right)
      ~,~~~~~\lambda_{\pm k} ~=~
        \pm {k\over R}\,
        \left(1 +  {m^2 R^2 \over k^2} - {m^4 R^4\over k^4} +...\right)~.
\label{tgdeg}
\eeq

Finally, it is also straightforward to explicitly solve for the light
mass eigenstates $|\tilde \nu_\pm\rangle$ corresponding to $k=0$.
To leading order in $mR$, we find
\beq
       |\tilde \nu_\pm \rangle ~=~ {1\over \sqrt{ 2}} \,
          \left\lbrace
         \left( 1- {\pi^2\over 6} m^2 R^2 \right)
        | \nu_L \rangle  ~\pm~ |\psi^{(0)}_1\rangle
            - mR \,\sum_{k=1}^\infty
         {1\over k}\,\left\lbrack |N^{(k)}\rangle -|M^{(k)}\rangle
                   \right\rbrack \right\rbrace~.
\label{eigenvec}
\eeq
This implies that the overlap between the light mass
eigenstates and the neutrino gauge eigenstate is generically less
than half in this scenario.
We will give an exact all-order solution for this eigenvector
in Sect.~3.

Thus, we conclude that in this orbifold case with
$m_N^{(n)}=m_M^{(n)}=m$ for all $n$,
all eigenvalues remain degenerate and there is no seesaw mechanism.

\subsection{Brane-shifting, lepton-number violation, and the induced seesaw}

We have already seen in Sect.~2.3 that a maximal seesaw emerges
in the case when $m_N^{(n)}$ and $m_M^{(m)}$ are unequal,
and likewise we have seen in Sect.~2.4 that the seesaw is
completely cancelled when $m_N^{(n)}$ and $m_M^{(m)}$ are precisely equal.
This suggests (and we shall shortly verify) that the magnitude of
the resulting seesaw is directly governed by the differences
$[m_N^{(n)}]^2 - [m_M^{(m)}]^2$.
It is therefore important, for the sake of our seesaw mechanism,
to generate such non-zero differences.  This is also important
if we want to split the ``light'' eigenvalues from each other
and thereby produce a Majorana
neutrino mass rather than a Dirac neutrino mass.

When the extra spacetime dimension is compactified on
a $\IZ_2$ orbifold, we have already seen in Sect.~2.2 that
it is natural for one of the two-component
right-handed spinors, \eg, $\psi_1$, to be taken to be even
under the $\IZ_2$ action $y\to -y$, while the other spinor
$\psi_2$ is taken to be odd.
If the left-handed neutrino $\nu_L$ is restricted
to a brane located at the orbifold fixed point $y=0$,
then $\psi_2$ vanishes at this point and so the most
natural coupling is between $\nu_L$ and $\psi_1$.
This then implies $m_N^{(n)}=m_M^{(n)}=m$ for all $n>0$.

It is therefore natural to ask under what conditions
the difference $[m_N^{(n)}]^2 - [m_M^{(m)}]^2$
can be non-zero.
Specifically, given that the straightforward orbifold coupling
gives a vanishing difference,
one wonders whether there might exist a physical
mechanism related to the orbifold that permits $m_N^{(n)}$ and
$m_M^{(n)}$ to be unequal.  Remarkably, however,
such a mechanism exists within Type~I string theory.
Specifically, in Type~I string theory, we have the freedom to
shift the branes away from the orbifold fixed points~\cite{shiftingrefs},
and under some restrictions, this shift can be done in a continuous way.
This then permits couplings between $\nu_L$ and more general
combinations of $\psi_1$ and $\psi_2$.
Thus, we see that brane-shifting provides us with a uniquely
``stringy'' method of breaking lepton number and generating a
seesaw mechanism and corresponding Majorana neutrino mass.\footnote{
      One might still wonder whether a coupling between $\nu_L$ and $\psi_2$
       can exist, given that such a term would not have been invariant under
the
      original $\IZ_2$ orbifold action.  However, it is improper to impose the
       orbifold projection on the spectrum and interactions corresponding
        to states on those branes that have been shifted away from the
          orbifold fixed points.  Instead, the orbifold action simply relates
          the wavefunctions of states on such a shifted brane to the
wavefunctions
          of states on another, oppositely shifted brane.  Thus, while
          the total theory still remains invariant under the orbifold
projection,
         the states on the individual shifted brane need not.}

Let us now analyze this situation in more detail.
Once again, we shall restrict our attention to the case of
five dimensions,
and imagine that the left-handed neutrino field $\nu_L$ (along
with the other Standard-Model fields) is restricted to a brane
whose bulk coordinate $y$ is shifted away from the orbifold fixed
point location $y=0$ to a general bulk coordinate $y^\ast$.
In such a situation, our generalized coupling between $\nu_L$
and the higher-dimensional $\Psi$ field takes the form
${\hat m}\bar\nu_L (\Psi+\Psi^{\rm c})|_{y*} + $ h.c.
Decomposing this into two-component spinors
and performing the Kaluza-Klein reduction as in Sect.~2.2
then yields the coupling
\beqn
       &&m \nu_L \,\left\lbrace
       \psi_1^{(0)} ~+~ \sqrt{2} \sum_{n=1}^\infty
      \left\lbrack \cos(n y^\ast/R) \psi_1^{(n)}
        + \sin(n y^\ast/R) \psi_2^{(n)}\right\rbrack
             \right\rbrace ~+~{\rm h.c.}\nonumber\\
      && ~~~=~  m \nu_L \psi_1^{(0)}
           ~+~ \sum_{n=1}^\infty  m_N^{(n)} \nu_L N^{(n)}
           ~+~ \sum_{n=1}^\infty  m_M^{(n)} \nu_L M^{(n)} ~+~ {\rm h.c.}
\label{shift}
\eeqn
where $m$ is defined in (\ref{mdef}) and where
\beqn
 m_N^{(n)} &\equiv&  m \,\left\lbrack
                \cos(ny^\ast/R) + \sin(ny^\ast/R)\right\rbrack\nonumber\\
 m_M^{(n)} &\equiv&  m \,\left\lbrack
                \cos(ny^\ast/R) - \sin(ny^\ast/R)\right\rbrack~.
\label{shiftedmdefs}
\eeqn
We thus see that brane-shifting has eliminated the degeneracy between
$m_N^{(n)}$ and $m_M^{(n)}$.

As before, the effect of this generalized coupling on the mass matrix
can be analyzed most conveniently by
integrating out all the massive Kaluza-Klein states.
This then gives rise to eigenvalues which again take the form
(\ref{tgeigpoly}),
except with $\mu$ now given by
\beq
   \mu  ~=~ -m^2 R \sum_{k=1}^n
           \,{1\over k}\, \sin\left({2 k y^\ast\over R}\right) ~.
\label{Meffshifted}
\eeq
Thus, we see that brane-shifting has succeeded in generating an effective
seesaw
between $\nu_L$ and $\psi_1^{(0)}$.

If the brane is at the orbifold fixed points $y^\ast=0$ or $y^\ast=\pi R$,
then $\mu=0$ and we recover the previous result (\ref{tgdeg})
for which the eigenvalues are degenerate.
In this case, the eigenvalues are suppressed only by the wavefunction
volume renormalization factors implicit in $m$.
This also is the result obtained at the midpoint $y^\ast=\pi R/2$.
However, at non-trivial values of $y^\ast$,
we find that the eigenvalues are no longer degenerate, and
instead experience a seesaw whose magnitude depends non-trivially
on the value of $y^\ast$.

\subsection{Bare Majorana masses:~ The case $M_0\not= 0$}

Let us now return to (\ref{newmatrix}), and
consider the case in which we include a
fundamental higher-dimensional Majorana-type mass term into our analysis.
As we discussed earlier, such lepton-number violating terms can be
generated
via {\bf 126}\/-type representations (when they arise), or
through other effective non-renormalizable terms that may appear
in the low-energy superpotential derived from a given string model.
Given that both the GUT scale and the string scale
can be reduced to the TeV-range in Type~I
string models with large extra dimensions~\cite{DDG},
it is natural to imagine that $M_0$ is close to the Type~I string scale,
which implies that $M_0\gg 1/R$.  However, we shall not make any
approximations based on this assumption in what follows.

In order to exhibit the effect of such a bare Majorana
mass on the resulting eigenvalues, we shall for simplicity
consider the orbifold case with $m=m_N^{(n)}=m_M^{(n)}$ for all $n$.
This then results in the mass matrix
\beq
      \calM ~=~ \pmatrix{
         0 &  m   &   m &   m  &   m  &  m & \ldots \cr
         m &  M_0 &   0  &   0  &   0  &  0  & \ldots \cr
         m &  0   &   M_0+1/R  &   0  &   0  &  0  & \ldots \cr
         m &  0   &   0  &   M_0-1/R  &   0  &  0  & \ldots \cr
         m &  0   &   0  &   0  &   M_0+2/R  &  0  & \ldots \cr
         m &  0   &   0  &   0  &   0   &  M_0-2/R  & \ldots \cr
         \vdots  &  \vdots &   \vdots  &   \vdots &   \vdots &  \vdots  &
\ddots \cr}~.
\label{Majmatrix}
\eeq
At this stage, however, it proves useful to define
\beq
         k_0 ~\equiv~ \lbrack M_0 R\rbrack ~,~~~~~
         \epsilon ~\equiv~ M_0 - {k_0\over R}~
\label{newvariables}
\eeq
where $[x]$ denotes the integer nearest to $x$.
Thus, $\epsilon$ is the smallest
diagonal entry in the mass matrix
(\ref{Majmatrix}), corresponding to the excited Kaluza-Klein state
$M^{(k_0)}$.  In other words, we have $\epsilon \equiv M_0$ (modulo $R^{-1}$),
satisfying $-\half R^{-1} < \epsilon \leq \half R^{-1}$.
The remaining diagonal entries in the mass matrix
can then be expressed as $\epsilon \pm k'/R$ where $k'\in \IZ^+$.
Upon suitably reordering the rows and columns of our mass matrix,
we can therefore cast this matrix into the form
\beq
      \calM ~=~ \pmatrix{
         0 &  m   &   m &   m  &   m  &  m & \ldots \cr
         m &  \eps &   0  &   0  &   0  &  0  & \ldots \cr
         m &  0   &   \eps+1/R  &   0  &   0  &  0  & \ldots \cr
         m &  0   &   0  &   \eps-1/R  &   0  &  0  & \ldots \cr
         m &  0   &   0  &   0  &   \eps+2/R  &  0  & \ldots \cr
         m &  0   &   0  &   0  &   0   &  \eps-2/R  & \ldots \cr
         \vdots  &  \vdots &   \vdots  &   \vdots &   \vdots &  \vdots  &
\ddots \cr}~.
\label{newMajmatrix}
\eeq

While this may look similar to our original mass matrix (\ref{Majmatrix}),
the important consequence of this rearrangement is that the {\it heavy}\/ mass
scale $M_0$ has been replaced by the {\it light}\/ mass scale $\epsilon$.
Unlike $M_0$, we see that $|\epsilon|\lsim{\cal O}(R^{-1})$.
Thus, the heavy Majorana mass scale $M_0$ completely {\it decouples}\/
from the physics!  Indeed, the value of $M_0$ enters the results only
through its determinations
of $k_0$ and the precise value of $\epsilon$.
At first sight, this may seem counter-intuitive,
because the (heavy) bare Majorana
mass $M_0$ would na\"\i vely appear to shift the
ground-state energy of the Kaluza-Klein tower, and thereby induce a
strong suppression for the neutrino mass.
However, as we have seen, the presence of the infinite
tower of regularly-spaced Kaluza-Klein states ensures
that only the value of $M_0$ modulo $R^{-1}$
plays a role.

The easiest way to solve (\ref{newMajmatrix}) for
the eigenvalues $\lambda_{\pm}$ is to
use the same diagram as in Fig.~\ref{newfigure},
and integrate out the Kaluza-Klein modes.
It turns out that there are two cases to consider,
depending on the value of $\epsilon$.  If $|\epsilon|  \gg m$
(which can arise when $mR\ll 1$),
then {\it all}\/ of the Kaluza-Klein modes are extremely
massive relative to $m$,
and we can integrate them out to obtain an effective
$\nu_L \nu_L$ mass term of size\footnote{
       In (\ref{cotresult}), care needs to be taken
       with respect to the order in which the terms
       are introduced into the (apparently divergent) Kaluza-Klein
       summations.  The only physically
       consistent organization of the terms that respects
       the symmetries of the Kaluza-Klein theory is to
       pair positive Kaluza-Klein modes with their corresponding
       negative Kaluza-Klein modes (or equivalently to pair the modes
       $N^{(k)}$  and $M^{(k)}$, which originally resulted from the
       algebraic decomposition of the fields $\psi_1$ and $\psi_2$).
       This pairing is therefore utilized in (\ref{cotresult}).
       It may seem at first that the reorganization of the
       Kaluza-Klein modes in passing from (\ref{Majmatrix}) to
       (\ref{newMajmatrix}) would render (\ref{cotresult}) invalid.
       However, if we were to directly integrate out the modes
       in the original order corresponding to (\ref{Majmatrix}),
       we would obtain the same expression as (\ref{cotresult})
       except with $\epsilon$ replaced by $M_0$.
       Because $M_0$ appears as the argument of a cotangent function
       in the result, this replacement is inconsequential.
       Thus, (\ref{cotresult}) remains correct as written.
       This conclusion will also hold for (\ref{secondcotresult}).
       Indeed, in general, we are free to reorder
       the rows and columns of this matrix without introducing any
       subtleties into the determination of the resulting eigenvalues.
       As a separate matter, we also note that integrating out the Kaluza-Klein
        states results in an overall sign
        which is opposite to that given in (\ref{cotresult}).
          However, interpreting this result as the physical neutrino mass
           allows us to disregard this sign.  }
\beqn
   |\epsilon|  \gg m:~~~~~~~~
      m_\nu &=&  m^2/\epsilon ~+~ m^2 \,\sum_{k'=1}^\infty
         \left( {1\over \epsilon+k'/R} + {1\over \epsilon-k'/R}
\right)\nonumber\\
         &=& \pi m^2 R \cot\left(\pi R\epsilon\right)~.
\label{cotresult}
\eeqn
We shall discuss the special case $\epsilon= \half R^{-1}$ in Sect.~4.
Alternatively, if $|\epsilon| \not\gg m $, then the lightest Kaluza-Klein
mode $M^{(k_0)}$ should not be integrated out, and we obtain an effective
$\nu_L \nu_L$ mass term of size $\mu$, where
\beqn
   |\epsilon| \not\gg m:~~~~~~~~
       \mu &\equiv& -m^2 \,\sum_{k'=1}^\infty
         \left( {1\over \epsilon+k'/R} + {1\over \epsilon-k'/R}
\right)\nonumber\\
         &=& {m^2\over \epsilon} - \pi m^2 R \cot\left(\pi R\epsilon\right)~.
\label{secondcotresult}
\eeqn
Note that $\mu \to 0$ smoothly as $\epsilon\to 0$, with $\mu$ otherwise
of size ${\cal O}(m^2 R)$.
Diagonalizing the final resulting $2\times 2$ mass matrix
between $\nu_L$ and $M^{(k_0)}$
in the presence of this mass term then yields the result
\beq
   |\epsilon|  \not\gg m:~~~~~~~~
   \lambda_\pm ~=~ \half \left\lbrack
     (\mu+\epsilon) ~\pm~
          \sqrt{ (\mu-\epsilon)^2 + 4 m^2 }\right\rbrack~.
\eeq
Thus, as $M_0\to 0$ (or as $M_0\to n/R$ where $n\in \IZ$),
we see that $\epsilon,\mu \to 0$, and we recover
the eigenvalues given in (\ref{tgdeg}).

We therefore conclude that although we may have started with
a bare Majorana mass $M_0\gg R^{-1}$,
in all cases the final neutrino mass remains of order $m^2 R$.
Even though we might have expected
a neutrino mass of order $m^2/M_0$ from
the mixing between $\nu_L$ and the original zero-mode $\psi_1^{(0)}$,
the contribution $m^2/M_0$ from the
zero-mode is completely cancelled by the summation over
the Kaluza-Klein tower, with the
seesaw between
$\nu_L$ and $M^{(k_0)}$ becoming dominant instead.
It is this feature that causes the heavy scale $M_0$ to be effectively
replaced by the radius $R^{-1}$, so that once again
our effective seesaw scale
is $M_{\rm eff}\sim {\cal O}(R^{-1})$.
Thus, we see that in a theory in which the heavy lepton experiences large extra
dimensions, the radius --- and not the bare lepton mass $M_0$ with which
we started --- plays the role of the heavy scale in the seesaw mechanism.

\subsection{General case: ~$y^\ast\not=0$, $M_0\not=0$}

Finally, for completeness, we turn to the general case in which we
include the effects of brane-shifting and bare Majorana masses
simultaneously.  Thus, we take $y^\ast\not=0$ and $M_0\not=0$.
As before, we define $\epsilon$ as in (\ref{newvariables}),
and for simplicity we shall restrict our attention to the case
where $|\epsilon| \gg m$.  We can therefore integrate out all
Kaluza-Klein modes, obtaining
\beqn
      m_\nu &=&    {m^2\over M_0} ~+~ \sum_{k=1}^\infty \,\Biggl\lbrace
            {m^2 \,[\cos(ky^\ast/R) + \sin(k y^\ast/R)]^2 \over M_0 + k/R }
                \nonumber\\
        && ~~~~~~~~~~~~~~~~~~~~~~~~~ ~+~
            {m^2 \,[\cos(ky^\ast/R) - \sin(k y^\ast/R)]^2 \over M_0 - k/R }
            \Biggr\rbrace \nonumber\\
       &=&  \pi m^2 R \left\lbrace
               \cot(\pi R\epsilon) + {\sin\left[ (\pi - 2y^\ast/R)  M_0
R\right] \over
                    \sin(\pi  M_0 R) } \right\rbrace ~.
\label{genresult}
\eeqn
 From this result, we see that when
brane-shifting and bare Majorana masses
are present {\it simultaneously}\/,
the bare Majorana mass $M_0$ does not completely decouple from the final
result in favor of $\epsilon$.  Instead, a slight dependence on $M_0$ remains
in the second term, due to the effects of the shifted brane.
However, once again this effect is small (appearing only within
the arguments of trigonometric functions), and the overall scale remains
$m_\nu \sim {\cal O}(m^2 R)$.  Thus, the effective seesaw scale in
this case is again $M_{\rm eff}\sim {\cal O}(R^{-1})$
rather than $M_0$.

\section{Higher-dimensional neutrino oscillations}
\setcounter{footnote}{0}

Let us now consider the implications of the above higher-dimensional
seesaw scenarios for neutrino oscillations.
Once again, we shall find that significant differences exist
relative to the usual four-dimensional case.

\subsection{Four-dimensional neutrino oscillations}

Let us begin by recalling how neutrino oscillations arise in
the usual four-dimensional case.
We suppose, in all generality, that we have two sets
of neutrinos,
a set of gauge eigenstates $\nu_f$
and a set of mass eigenstates $\tilde \nu_i$
which are non-trivially
related to each other through a unitary mixing matrix $U$:
\beq
               \nu_f ~=~ \sum_i \,U_{fi}\, \tilde\nu_i~.
\eeq
Here the tilde indicates a mass eigenstate.
This matrix $U$ is uniquely determined from the mass mixing
matrix $\calM$ involved in the seesaw mechanism, and is nothing
but the inverse of the matrix of eigenvectors of $\calM$.
It is the fact that the gauge eigenstates are non-trivial
combinations of the physical propagating mass eigenstates
that causes the gauge eigenstates to oscillate as a function
of time.
Specifically, given such a mixing matrix $U$,
we find that the probability of oscillation from
$\nu_f$ to $\nu_{f'}$
after time $t$ is given by
\beq
 P_{f\to f'}(t) ~=~ \sum_i \,\left| U_{fi} U_{f'i}\right|^2 ~+~
            2\,\sum_{i > j}\,
         {\rm Re}\, \left\lbrace
        U_{fi}U^\ast_{f'i} U^\ast_{fj} U_{f'j}
         \, \exp\left\lbrack  {i (E_j - E_i) t}
           \right\rbrack \right\rbrace~
\label{genform}
\eeq
where $E_i\equiv (p^2 + m_i^2)^{1/2}$ is the energy of $\tilde \nu_i$.
In the extreme relativistic limit for which we assume that all
neutrinos have the same momentum $p\gg m_i$,
we can approximate $E_j-E_i \approx (m_j^2 -m_i^2)/2p$.
For $f\not= f'$, we thus see that
this probability can be non-zero only if $m_i\not= m_j$
for some pair of mass eigenstates $(i,j)$ for which the appropriate
matrix elements of $U$ are non-vanishing.

Of particular interest is the total probability that a given neutrino
$\nu_f$ oscillates into {\it any}\/ other state.  This deficit probability
is the complement of the probability that this neutrino
is preserved, and from (\ref{genform}) this
preservation probability can be easily evaluated as
\beq
      P_{f\to f}(t) ~=~ \biggl| \sum_i \left|U_{fi}\right|^2
         \exp(i E_i t)\biggr|^2~.
\label{deficit}
\eeq

Note that these results apply
not only to flavor oscillations (in which case we interpret the $f$ index as
indicating flavor), but also to neutrino/anti-neutrino oscillations
(in which case we identify $\nu_f = (\nu,N)$ for a fixed flavor)
as well as general combinations of the two.

\subsection{Higher-dimensional neutrino oscillations}

The above formalism carries over directly into the higher-dimensional
seesaw scenarios we presented in Sect.~2.
For simplicity, let us focus first on
neutrino/anti-neutrino oscillations, and consider the
orbifold case discussed in Sect.~2.4.
We have seen in Sect.~2 that in higher dimensions,
it is natural
to imagine a Kaluza-Klein tower for the right-handed neutrino, and that this
automatically leads to a mixing mass matrix of the form (\ref{tgnewmatrix}).
This then generates a set of corresponding mass eigenstates which
we can denote
\beq
        \tilde \calN^T ~\equiv~ (\tilde \nu_L, \tilde \psi_1^{(0)},
                  \tilde N^{(1)}, \tilde M^{(1)},
                  \tilde N^{(2)}, \tilde M^{(2)}, ...)~
\label{tildecalNdefnew}
\eeq
in analogy with (\ref{calNdef}).
Given these results, we see that (\ref{genform}) and (\ref{deficit})
continue to hold;
as before, we simply identify the matrix $U$ in (\ref{genform}) as
the inverse of the matrix of eigenvectors
of the mass matrix $\calM$ given in (\ref{tgnewmatrix}).
Specifically, we write
\beq
       \calN ~=~ U \tilde \calN
\eeq
where the gauge eigenstates $\calN$ are defined in (\ref{calNdef})
and the mass eigenstates $\tilde \calN$ are defined in (\ref{tildecalNdefnew}).
Note that since $U$ is (by definition)
the matrix that diagonalizes $\calM$,
the non-diagonality of $\calM$ implies the non-diagonality of $U$.

It turns out to be remarkably simple to obtain an {\it exact}\/ result for
this $U$-matrix in the case (\ref{tgnewmatrix}), valid
for all values of $mR$.
We find that $U$ is given by
\beq
          U^\dagger ~=~ \pmatrix{
                {\bf U}_+ \cr
                {\bf U}_- \cr
                {\bf U}_{+1} \cr
                {\bf U}_{-1} \cr
                {\bf U}_{+2} \cr
                {\bf U}_{-2} \cr
               \vdots \cr}
\label{Umatrixgenform}
\eeq
where each individual row (\ie, each individual eigenvector)
is given exactly by
\beq
           {\bf U}_i ~\equiv~ {1\over \sqrt{N_i}}\, \left(
             1,~ {m\over \lambda_i},~
                {m\over \lambda_i-1/R},~
                {m\over \lambda_i+1/R},~
                {m\over \lambda_i-2/R},~
                {m\over \lambda_i+2/R},~ ...\right)~.
\label{row}
\eeq
Here $\lambda_i$ are the mass eigenvalues
which are the exact solutions to (\ref{trans1}),
and likewise the normalization factor $N_i$ in (\ref{row})
is given exactly by
\beq
            N_i ~=~ 1 + {\pi^2 m^2 R^2 \over \sin^2(\pi \lambda_i R)} ~=~
               1 + \pi^2 m^2 R^2 + {\lambda_i^2\over m^2}~,
\eeq
where we have used (\ref{trans1}) in the final equality.
Note that approximate expressions for the eigenvalues $\lambda_i$
are given for $mR\ll 1$ in (\ref{gensoln}) and (\ref{tgdeg}).
Given the result (\ref{Umatrixgenform}) for the $U$-matrix,
it is then straightforward to show that
\beq
              U^\dagger \calM U ~=~ {\rm
diag}(\lambda_+,\lambda_-,\lambda_{+1},
               \lambda_{-1},\lambda_{+2},\lambda_{-2},...)~,
\eeq
thereby verifying that this $U$-matrix indeed diagonalizes $\calM$.

\begin{figure}[ht]
\centerline{
      \epsfxsize 3.1 truein \epsfbox {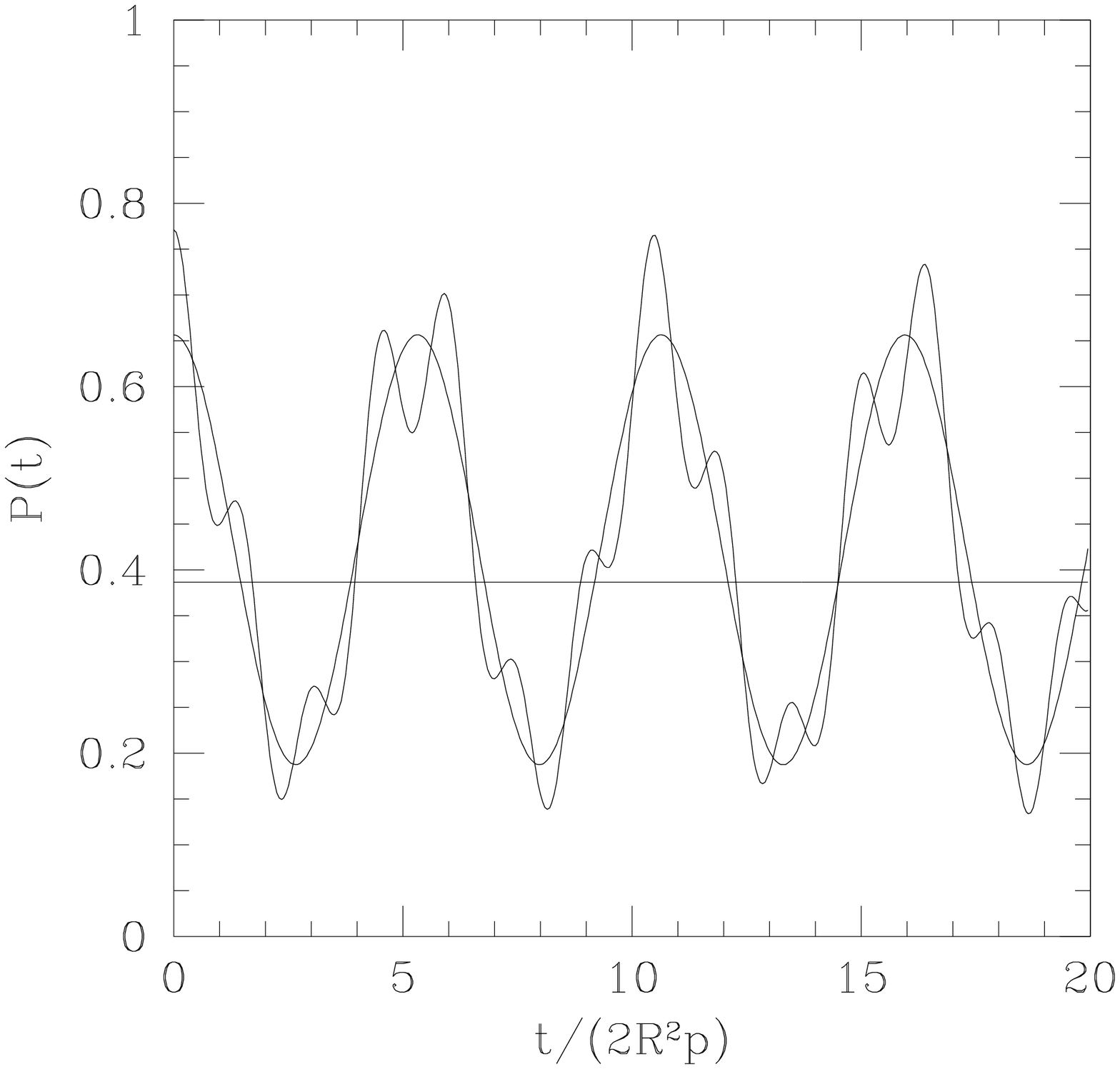}
      \hskip 0.2 truein
      \epsfxsize 3.1 truein \epsfbox {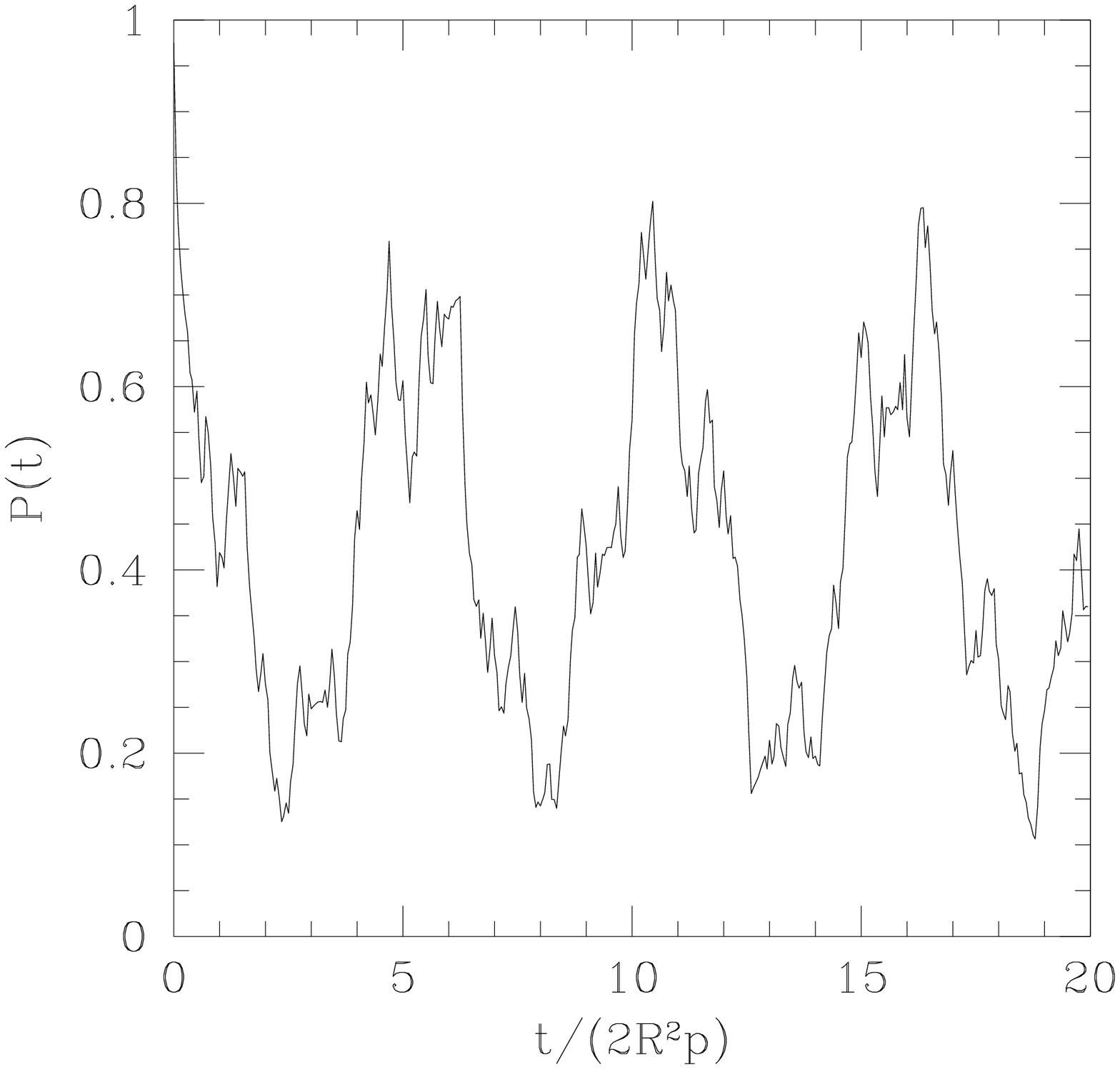}}
\caption{
     Higher-dimensional neutrino oscillations in
     the orbifold scenario discussed in Sect.~2.4.
      (a)  The evolution of the probability sum in (\ref{deficit2})
               as more and more Kaluza-Klein states are
            included in the sum.
           We have taken $mR=0.4$.
         The flat line shows the contribution when only
         the degenerate zero-mode $\lambda_\pm$ eigenvalues are included
          (no oscillations);
        the cosine shows the probability when
         the first excited Kaluza-Klein
         states are also included;  and the irregular curve shows
          the interference that results when the
       second excited Kaluza-Klein states are also included.
       Note that the initial probability $P(t=0)$ approaches $1$
        as the full spectrum of Kaluza-Klein states is included.
      (b)  The final result:  the total probability that the
     gauge neutrino $\nu_L$ is preserved as a function of time
      when all Kaluza-Klein states are included.
     The multi-component nature of
     the neutrino oscillation is reflected in the
     jagged shape of the oscillations,
     as well as in the fact that the resulting neutrino deficits and
     regenerations, though sizable, are never total.}
\label{oscfig1}
\end{figure}

Given this $U$-matrix, is straightforward to calculate
the corresponding probability of the neutrino
gauge eigenstate $\nu_L$ oscillating into
any of the Kaluza-Klein excited states
$\lbrace\psi_1^{(0)}, N^{(k)}, M^{(k)}\rbrace$, or conversely
the probability that the neutrino $\nu_L$ is preserved as a function
of time.
Using (\ref{deficit}) and (\ref{row}), we see that the latter
probability is
simply given by\footnote{In writing (\ref{deficit2}), we
       have assumed that the relativistic approximation $p \gg m_i$
       continues to hold for the
       excited Kaluza-Klein states.   Of course, at a formal level,
       when the summation is taken to infinity we eventually
       exceed this bound and the extremely heavy Kaluza-Klein states
       become non-relativistic.  Therefore, for such states, the relativistic
       approximation should no longer be used.  This will have
       little practical consequence, however, since the contributions
       to the oscillation probability from the extremely heavy Kaluza-Klein
       states are strongly suppressed.}
\beq
        P_{\nu_L\to\nu_L}(t) ~=~ \biggl | \sum_{i}  {1\over N_i}\,
         \exp\left( {i \lambda_i^2 t\over 2 p}\right)\biggr|^2~.
\label{deficit2}
\eeq
The result is plotted in Fig.~\ref{oscfig1}.

One possible worry regarding such higher-dimensional oscillation scenarios
might have initially seemed to be
that because the neutrino oscillations take place
within an infinite-state system whose mass eigenvalues are not
commensurate, the resulting oscillations would tend to
interfere destructively,
thereby amounting to a neutrino ``damping'' without any
possibility of neutrino regeneration.
However, we now see from Fig.~\ref{oscfig1} that this is not the case,
and we indeed continue to have oscillations with both neutrino
deficits and neutrino regeneration.
Thus, we see that while the multi-state oscillation has eliminated
the formal {\it periodicity}\/ of the oscillation probability as
a function of time, the final result is still effectively periodic.
This is the reflection of the fact that the dominant component of
the oscillation is the simple two-state oscillation between
the zero-mode neutrino states and the first excited Kaluza-Klein states.
The ``wavelength'' of this oscillation is thus set by the lowest-lying
eigenvalue difference $\lambda^2_{\pm 1}- \lambda_\pm^2$.
However, the striking
signature of the multi-state
nature of the oscillations is (in addition to
their jagged profile) the fact that the neutrino
never {\it completely}\/ oscillates away
or is restored --- \ie, the neutrino
deficits and regenerations are never total.  This is therefore
in strong contrast to the simpler case of two-component oscillations.

This result is not qualitatively affected when we consider
the cases involving brane-shifting or bare Majorana masses.
In the brane-shifted case,
the result (\ref{row}) continues to hold, where $m$ is
replaced by $m_N^{(k)}$ for the components corresponding to
$N^{(k)}$
and by
$m_M^{(k)}$ for the components corresponding to $M^{(k)}$.
Finally, we can also consider the case when a bare Majorana mass $M_0$
is included.
In such situations, we find that
the oscillation into the lowest Kaluza-Klein state
is proportional to $m^2/\epsilon^2$, where
$\epsilon\equiv M_0$ (modulo $R^{-1}$).

It is important to stress that
there is also the usual possibility of flavor oscillations
in addition to the oscillations into Kaluza-Klein states
that we are discussing here.
However, the above sorts of higher-dimensional oscillation scenarios
easily generalize to the case of flavor oscillations:
we simply introduce an additional flavor index, and imagine that
our mass matrices are also non-diagonal in flavor space.
Note that
this last assumption is completely analogous to what must be assumed
in the ordinary four-dimensional case.  We then likewise find that
the above neutrino/anti-neutrino oscillations can also indirectly induce
flavor oscillations, with
the flavor oscillations
occurring indirectly through the masses and flavor mixings of the
corresponding excited Kaluza-Klein states.
In fact,
such indirect flavor oscillations through excited Kaluza-Klein
states can be viewed as the higher-dimensional analogue of the
indirect flavor oscillations discussed in Ref.~\cite{indirect}.

\subsection{Comparison with experiment}

Given the above results, the recent experimental detection
of neutrino oscillations can be used to estimate the level spacings
of the Kaluza-Klein states, which in turn permits
us to estimate
the size of the radius required.  In the normal four-dimensional
scenario, a neutrino mass difference of the order
$\delta m^2\sim 10^{-4}$ eV$^2$ is quoted~\cite{superK} as being sufficient
to explain the oscillation observed at SuperKamiokande.
In our higher-dimensional
scenario, however, we have seen that
this mass difference can be attributed not to
the left-handed neutrinos, but to the
Kaluza-Klein tower of $\tilde N^{(n)}$ and $\tilde M^{(n)}$ mass eigenstates
whose masses are given by the eigenvalues
$\lbrace\lambda_\pm, \lambda_{\pm k}\rbrace$, $k\in \IZ^+$.
We have seen that for $mR\ll 1$, these eigenvalues are approximately
given by $\lambda_{\pm k}\approx k/R$;  we have also seen above that
the effective oscillation ``wavelength'' is set by the lowest-lying
eigenvalue difference $\lambda^2_{\pm 1} - \lambda^2_\pm\approx R^{-2}$.
Thus, we can roughly associate
$\delta m^2$
with $R^{-2}$, obtaining the estimate $R\approx 10^{-5}$ meters.
Such an
extra dimension would therefore be perfectly consistent with the
scenario advocated in Ref.~\cite{Dim}, which would in turn enable us
to identify the
extra dimension we have been discussing as one which only gravity
(and our higher-dimensional $\Psi$ field) can experience.
As discussed in Ref.~\cite{Dim}, a ``gravity-only'' extra dimension
of this size is believed to be consistent with all laboratory,
astrophysical, and cosmological constraints.

Pursuing this line of reasoning a bit further,
we may even use the results given in (\ref{mRvalue}) and (\ref{deficit})
in conjunction with the mixing-parameter bound $\sin^2 2\theta >0.82$
given in Ref.~\cite{superK}.  If we associate the finite value of $n$
with $M_s R$ (as might be expected in an effective field-theory
approach where we keep only the lowest excitations of the Kaluza-Klein
tower), we can obtain a rough
bound on the string scale $M_s \lsim \hat m \sim 1$~TeV for
$\delta =2$ and for a Yukawa coupling $\sim {\cal O}(1)$ for oscillations into
the first Kaluza-Klein state.
Thus, it would appear that the experimental bound on the mixing
rules out larger values of $\delta$, so that only
two extra dimensions felt by the higher-dimensional neutrino field are
consistent
with the SuperKamiokande results.
However, more flexibility is allowed in the case of flavor oscillations,
which we have not discussed here.

Of course, the above analysis
is at best only qualitative.
Ultimately, one would also like to take into account the data
concerning both atmospheric and solar neutrinos.
Likewise, one would one would
also need to take into account the energy-dependence of the
experimental signals.
We leave this subject for future investigation.

\section{Neutrino oscillations without neutrino masses}
\setcounter{footnote}{0}

Finally, let us turn our attention to something
far more speculative:  the possibility of neutrino oscillations
 {\it without}\/ neutrino masses.

In the usual four-dimensional seesaw mechanism,
neutrino oscillations require (and therefore can be interpreted as
the unique signature of) neutrino masses.
Let us recall why this is the case for
neutrino/anti-neutrino oscillations.
In the usual seesaw mechanism, we are required to have a
mass matrix of the form (\ref{usualmatrix}).  Regardless of the
Majorana mass $M$ of the right-handed neutrino $N$,
the only way to achieve a massless neutrino in
this scenario is to set $m=0$.  However, this then results in
a diagonal mass matrix, so that the corresponding matrix $U$ of eigenvectors
is also diagonal.  Therefore no oscillations are produced.
Consequently, the only way to have neutrino oscillations in this scenario
is to have neutrino masses.  This argument can also be
extended to the case of flavor oscillations.

The crucial ingredient in the above four-dimensional argument
is that matrices of the form (\ref{usualmatrix})
cannot have a vanishing mass eigenvalue without
being diagonal.
In higher dimensions, however, we have seen that our mass
matrices are infinite-dimensional (corresponding to mixings
between the infinite numbers of Kaluza-Klein modes), and therefore
this constraint may be relaxed.
This then would permit the possibility of neutrino oscillations
without neutrino masses.

As a concrete example of this phenomenon, let us
consider the results of Sect.~2.6, where
we examined the consequences of introducing a bare Majorana mass
$M_0$ for the higher-dimensional field $\Psi$.
We showed, remarkably, that the overall scale of this Majorana mass
completely decouples from the problem, and that only
$\epsilon \equiv M_0$ (mod $R^{-1}$)
plays a role in determining the mass of the resulting
neutrino mass eigenstate.
Let us now consider what happens in the special case
that $\epsilon= \half R^{-1}$.
After further reordering of the rows and columns
corresponding to the excited Kaluza-Klein states,
the mass matrix (\ref{newMajmatrix}) then takes the form
\beq
      \calM ~=~ \pmatrix{
         0 &  m   &   m &   m  &   m  &  m & m & \ldots \cr
         m &  1/(2R) &   0  &   0  &   0  &  0  & 0 & \ldots \cr
         m &  0   &   -1/(2R)  &   0  &   0  &  0  & 0 & \ldots \cr
         m &  0   &   0  &   3/(2R)  &   0  &  0  & 0 & \ldots \cr
         m &  0   &   0  &   0  &   -3/(2R)  &  0  & 0 & \ldots \cr
         m &  0   &   0  &   0  &   0   &  5/(2R)  & 0 & \ldots \cr
         m &  0   &   0  &   0  &   0   &  0  & -5/(2R) & \ldots \cr
         \vdots  &  \vdots &   \vdots  &   \vdots &
         \vdots &  \vdots  & \vdots & \ddots \cr}~.
\label{newnewMajmatrix}
\eeq
In order to obtain the corresponding neutrino mass,
we note that for $\epsilon=\half R^{-1}$, the assumption $mR\ll 1$
translates into $\epsilon\gg m$, whereupon the result (\ref{cotresult})
is valid.  Thus, for $\epsilon=\half R^{-1}$, we find the remarkable
result that $m_\nu =0$!
In obtaining this result, one might worry that (\ref{cotresult}) is
only approximate because it relies on the procedure of integrating
out the Kaluza-Klein states rather than a full diagonalization of
the corresponding mass matrix.  However, it is straightforward to show
that when $\epsilon=\half R^{-1}$, the characteristic eigenvalue
equation $\det (\calM-\lambda I)=0$ for the mass matrix
(\ref{newnewMajmatrix}) becomes
\beq
      \lambda R
       \left\lbrack
     \prod_{k=1}^\infty
            (\lambda^2 R^2 - (k -\half)^2) \right\rbrack
        \left\lbrack
       1 - 2 m^2 R^2 \sum_{k=1}^\infty {1\over  \lambda^2 R^2 - (k-1/2)^2 }
       \right\rbrack ~=~ 0~.
\label{charvanishing}
\eeq
This has an exact trivial solution $\lambda=0$, corresponding to an
exactly massless neutrino.
Thus, we conclude that $m_\nu=0$ for $\epsilon=\half R^{-1}$, regardless
of the relative sizes of $m$ and $R$.

\begin{figure}[t]
\centerline{
      \epsfxsize 4.0 truein \epsfbox {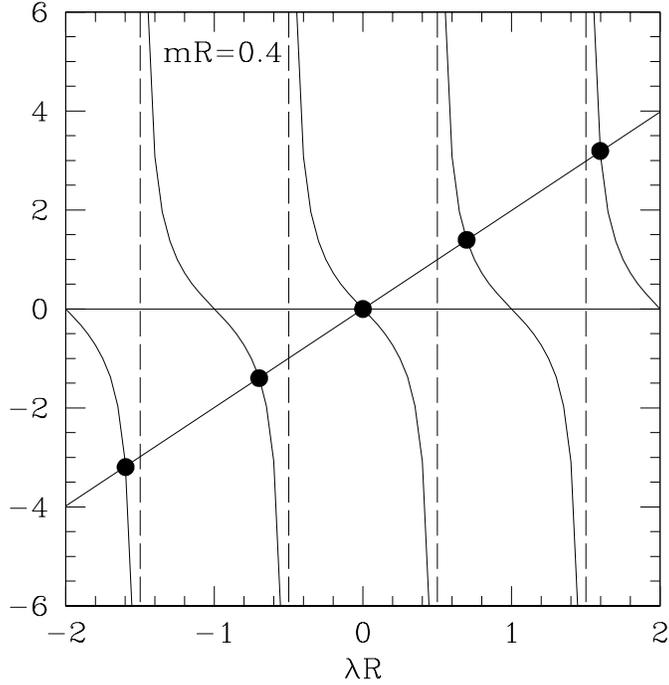}}
\caption{  Eigenvalue solutions to (\protect\ref{trans2}),
           represented as those values of $\lambda$ for which
          $-\tan(\pi \lambda R)$ intersects
         $\lambda R/ [\pi (mR)^2]$.  We have taken the
               fixed value $mR=0.4$ for this plot.
           The behavior of the eigenvalues as functions of $mR$
          can be determined graphically by changing the slope
           of the intersecting diagonal line.
         Regardless of the value of $mR$, we see that the zero
         eigenvalue is fixed and unique.}
\label{orbfigurenew}
\end{figure}

There is also another useful way to understand the emergence of
this vanishing eigenvalue.
For $m\not=0$,  it is straightforward to see that
the value $\lambda=(k+\half)/R$ is not a solution
of (\ref{charvanishing}) because the cancellation of the first bracketed
factor is offset by the divergence of the second bracketed factor.
We can therefore restrict our attention to the second bracketed factor,
and reduce (\ref{charvanishing}) to the form
\beq
       \lambda R ~=~ -\pi (mR)^2 \,\tan\left(\pi \lambda R\right)~.
\label{trans2}
\eeq
This equation is the analogue of (\ref{trans1}), and upon
plotting this condition graphically, we find the result shown
in Fig.~\ref{orbfigurenew}.
Remarkably, this is effectively the same as Fig.~\ref{orbfigures}(a)
except that the cotangent curves of Fig.~\ref{orbfigures}(a) have been
translated horizontally by $\half  \lambda R$.
We thus see that the effect of adding a bare Majorana mass
term corresponding to $\epsilon =\half R^{-1}$ is simply to
 {\it shift}\/ the positions of the cotangent curves by exactly
half of the oscillation period.
This explains graphically why the zero eigenvalue emerges
precisely in the case $\epsilon=\half R^{-1}$.
We also see from this figure that the zero eigenvalue is independent
of the value of $mR$, and is unique.

Note that this graphical result is completely general:
the effect of adding a general bare Majorana mass
term $M_0$ is simply to {\it shift}\/ the positions
of the cotangent curves by an amount proportional to $M_0$.
In fact, by changing the value of $M_0$, we see that
it is possible to smoothly {\it interpolate}\/ between
the orbifold scenario  discussed Sect.~2.4 and
the scenario we are discussing here.
This also provides another explanation of why only
the value $\epsilon\equiv M_0$ (modulo $R^{-1}$) is relevant physically.
The regular, repeating aspect of the infinite towers of Kaluza-Klein
states is now manifested graphically in the periodic nature of
the cotangent function.

We can also solve for the full spectrum of eigenvalues as a function
of $mR$.  Following the same steps as in Sect.~2.4, we find that the
non-zero eigenvalues are identical to those given in (\ref{gensoln})
for $k\not=0$, except with $k\to k-\half$.  To order ${\cal O}(m^5 R^5)$,
this yields the non-zero eigenvalues
\beq
        \lambda_{\pm k} ~=~ \pm {k-1/2 \over R}\,
        \left\lbrack
        1 +  {m^2 R^2 \over (k-1/2)^2} - {m^4 R^4\over (k-1/2)^4}
+...\right\rbrack~,
         ~~~~ k>0~.
\label{tgdeg2}
\eeq

It is also straightforward to calculate the exact neutrino mass eigenstate
that corresponds to our vanishing eigenvalue.
If we use the basis of gauge eigenstates corresponding to the mass
matrix (\ref{newnewMajmatrix}), and denote this basis as
$\lbrace \nu_L, \hat N^{(1)}, \hat M^{(1)}, \hat N^{(2)}, \hat M^{(2)},
...\rbrace$,
we easily see that the neutrino mass eigenstate $|\tilde \nu_L\rangle$ is given
by the (normalized) result
\beq
       |\tilde \nu_L\rangle ~=~ {1\over \sqrt{ 1+ \pi^2 m^2 R^2}} \,
          \left\lbrace
         | \nu_L \rangle  ~-~  mR \,\sum_{k=1}^\infty
         {1\over k-1/2}\,\left\lbrack |\hat N^{(k)}\rangle -|\hat
M^{(k)}\rangle
                   \right\rbrack \right\rbrace~.
\label{eigenvec2}
\eeq
This result is exact for all $mR$, and is essentially
the analogue of (\ref{eigenvec}) in which one replaces $k\to k-1/2$.
Also note that this neutrino mass eigenstate
is {\it primarily}\/ composed of the neutrino gauge
eigenstate $\nu_L$, since $mR\ll 1$.
Although this neutrino mass eigenstate also contains
a small, non-trivial admixture of Kaluza-Klein states, the dominant
component of our massless eigenstate is still the
gauge-eigenstate neutrino $\nu_L$, as required phenomenologically.
Nevertheless, this combined neutrino mass eigenstate is
exactly massless in the limit
that the full, infinite tower of Kaluza-Klein states participates
in the mixing!  We stress that this remarkable result is valid
 {\it regardless}\/ of the value of neutrino Yukawa coupling $m$ or
the radius scale $R^{-1}$ of the Kaluza-Klein states.

\begin{figure}[ht]
\centerline{
      \epsfxsize 3.1 truein \epsfbox {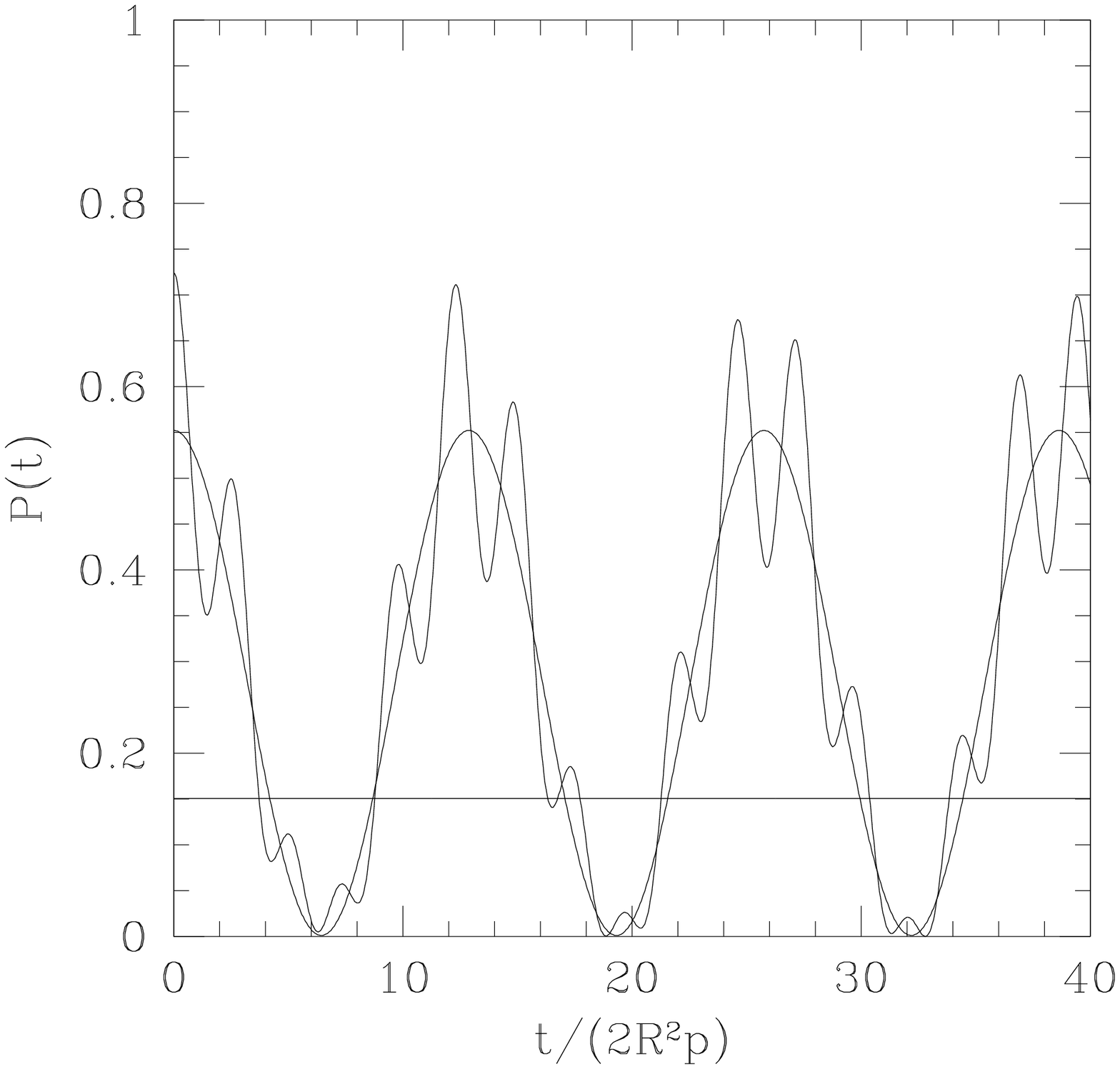}
      \hskip 0.2 truein
      \epsfxsize 3.1 truein \epsfbox {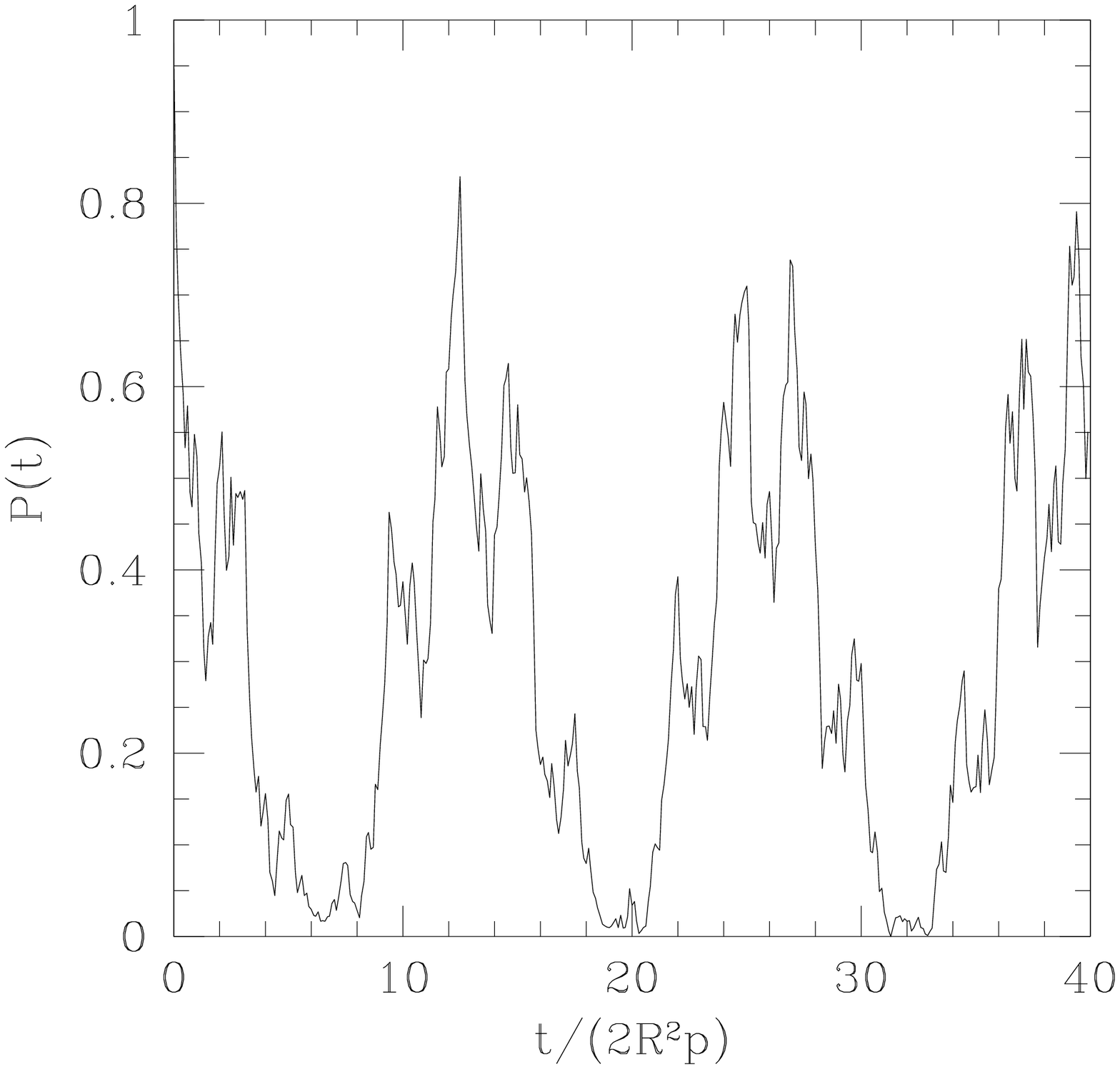}}
\caption{
     Higher-dimensional neutrino oscillations, even when the
     neutrino itself is massless.
      (a)  The evolution of the probability sum in (\ref{deficit2})
               as more and more Kaluza-Klein states are
            included in the sum.
           We have taken $mR=0.4$.
         The flat line shows the contribution when only
         the massless neutrino is included
          (no oscillations);
        the cosine shows the probability when
         the first excited Kaluza-Klein
         states are also included;  and the irregular curve shows
          the interference that results when the
       second excited Kaluza-Klein states are also included.
       Note that the initial probability $P(t=0)$ approaches $1$
        as the full spectrum of Kaluza-Klein states is included.
      (b)  The final result:  the total probability that the
     gauge neutrino $\nu_L$ is preserved as a function of time
      when all Kaluza-Klein states are included.
     The multi-component nature of
     the neutrino oscillation is reflected in
      the jagged shape of the oscillations.
     Unlike the oscillation in Fig.~\protect\ref{oscfig1}(b), however,
      in this case the deficits are total even though the regenerations
       are not.}
\label{oscfig2}
\end{figure}

Given this result,
it is straightforward to calculate the
probability that the neutrino gauge eigenstate oscillates into any of the
excited Kaluza-Klein states $\hat N^{(k)}$ or $\hat M^{(k)}$,
or conversely
the probability that the neutrino $\nu_L$ is preserved as a function
of time.
We find that the $U$-matrix corresponding to (\ref{newnewMajmatrix})
once again takes the form (\ref{Umatrixgenform}), where now the
individual rows (\ie, eigenvectors) are given exactly for all values of
$mR$ by
\beq
           {\bf U}_i ~\equiv~ {1\over \sqrt{N_i}}\, \left(
            1,~ {m\over \lambda_i-1/(2R)},~
                {m\over \lambda_i+1/(2R)},~
                {m\over \lambda_i-3/(2R)},~
                {m\over \lambda_i+3/(2R)},~ ...\right)
\label{newevectors}
\eeq
and the normalization factor $N_i$ in (\ref{newevectors}) is
given by
\beq
          N_i~=~ 1 + {\pi^2 m^2 R^2 \over \cos^2(\pi_i \lambda R)} ~=~
                1+ \pi^2 m^2 R^2 + {\lambda_i^2\over m^2}~.
\eeq
Here we have used (\ref{trans2}) in the final equality.
Substituting this result into (\ref{deficit}), we then find the
preservation probability shown in Fig.~\ref{oscfig2}.

Fig.~\ref{oscfig2} provides an explicit verification
that neutrino oscillations do indeed occur, even
though the physical neutrino is exactly massless.
Of course, this result is expected,
because the mass matrix (\ref{newnewMajmatrix}) is non-diagonal.
Therefore the full mixing matrix $U$ that diagonalizes $\calM$
must also be non-diagonal.  The fact that this $U$-matrix
is non-diagonal then leads to the non-trivial mixings that produce
oscillations.

Just as in Sect.~3,
we observe that these higher-dimensional neutrino oscillations
lead to neutrino deficits as well as neutrino regeneration,
in a roughly periodic manner.
Specifically, there is no neutrino ``damping'' arising from the separate
incommensurate Kaluza-Klein oscillations, as might have been feared.
Howeever, unlike the oscillation in Sect.~3,
we see that in this scenario the deficits are total
even though the regenerations are not.
This could therefore serve as a  potential
experimental method of distinguishing between this scenario
and that discussed in Sect.~2.4.

Thus, we conclude
that the neutrino mass eigenstate
oscillates into the entire
tower of higher-dimensional Kaluza-Klein neutrinos, even though
it has no mass of its own!
Indeed, the masses of the right-handed Kaluza-Klein states themselves
are sufficient
to generate the desired oscillations indirectly.
Although this mechanism applies for neutrino/anti-neutrino
oscillations, it can also easily be generalized to accommodate flavor
oscillations as well, even if $(\nu_e,\nu_\mu,\nu_\tau)$ are
all taken to be massless.

One possible drawback to this scenario might initially seem to
be that it requires the precise value $\epsilon=\half R^{-1}$,
corresponding to a precise bare Majorana mass of the form
$M_0= (n+\half)/R$ where $n\in \IZ$.
This would therefore seem to require a precise fine-tuning.
However, it turns out that bare Majorana masses of precisely this
form emerge naturally from Scherk-Schwarz decompositions in string theory.
Recall that our original five-dimensional Dirac spinor field $\Psi$
is decomposed in the Weyl basis as $\Psi=(\psi_1,\bar\psi_2)^T$,
where $\psi_1$ and $\psi_2$ individually have the orbifold
mode-expansions given in (\ref{KKdecomp}).
However, let us consider performing a local rotation in $(\psi_1,\psi_2)$
space of the form
\beq
         \pmatrix{ \hat \psi_1 \cr \hat \psi_2 } ~\equiv~  R
         \pmatrix{ \psi_1 \cr \psi_2 } ~~~~~~~{\rm where}~~~
     R ~\equiv ~ \pmatrix {
               \cos (\omega y/R) & -\sin (\omega y/R) \cr
               \sin (\omega y/R) & \cos (\omega y/R) \cr}~.
\label{scherkschwarz}
\eeq
Such a general rotation is allowed in field theory because it
corresponds to a $U(1)$ symmetry of the higher-dimensional theory.
However, in string theory there are additional topological
constraints (coming from the preservation of the form of the worldsheet
supercurrent) that permit only {\it discrete}\/ rotations.
In particular, in a compactification from five to four dimensions,
this restriction limits us to the only non-trivial
possibility $\omega=1/2$.
(The trivial case $\omega=0$ corresponds to the
straightforward orbifold situation discussed in Sect.~2.4.)
Taking $\omega=1/2$ then implies $\psi_{1,2}(2\pi R) = -\psi_{1,2}(0)$, which
shows that lepton number is broken globally (although not locally)
as the spinor is taken around the compactified space.
This is the result of the discrete ``twist'' induced by the Scherk-Schwarz
$R$-matrix.
After the Kaluza-Klein decomposition,
this breaking of lepton number in turn induces a
Majorana mass term with $M_0= \omega/R$ (modulo $R^{-1}$).
Thus, we see that exactly the desired value of the Majorana mass
emerges naturally from a Scherk-Schwarz decomposition, for reasons
that are {\it topological}\/ and hence do not require any fine-tuning.

Thus, in this respect, the scenario that leads to an exactly massless
neutrino is the Scherk-Schwarz ``twisted'' counterpart of the straightforward
orbifold scenario of Sect.~2.4.  Relative to the orbifold scenario,
we see that this twisting introduces
lepton-number violation in a natural way, brings the neutrino mass
to zero, and also breaks the
two-fold degeneracy for the lightest ground state --- all
while maintaining neutrino oscillations.

Note also that the masslessness of the neutrino mass eigenstate
relies rather crucially on taking the full $n\to\infty$ limit in
the above calculation.
This might seem to
go against the spirit of the effective field-theory approach
we have been following wherein we would truncate
the Kaluza-Klein sum at $n_{\rm max}\sim {\cal O}(M_s R)$.
Nevertheless, it is not unreasonable to expect that
in the full underlying string theory, a similar
mechanism might be implemented once all of the string states
(not only Kaluza-Klein states, but also winding states and
oscillator states) are properly included.
This would represent a uniquely ``stringy'' behavior,
not unlike the Hagedorn phenomenon~\cite{Hagedorn} which
also emerges only when {\it all}\/ string states are included.

Indeed, even within the effective field-theory approach that
we have been following, the resulting neutrino mass is
extraordinarily suppressed.
To see this, let us imagine truncating our Kaluza-Klein levels
at $n_{\rm max}\sim {\cal O}(M_s R)$ where $M_s$ is the mass
scale of the underlying (string) theory, and let us take $M_0=\half R^{-1}$
as suggested by the Scherk-Schwarz analysis above.
In this case, the original mass matrix (\ref{Majmatrix}) takes
the form
\beq
      \calM ~=~ \pmatrix{
         0 &  m   &   m &   m  &   m  &  m &  \ldots \cr
         m &  1/(2R) &   0  &   0  &   0  &  0  &  \ldots \cr
         m &  0   &   3/(2R)  &   0  &   0  &  0  &  \ldots \cr
         m &  0   &   0  &   -1/(2R)  &   0  &  0  &  \ldots \cr
         m &  0   &   0  &   0  &   5/(2R)  &  0  &  \ldots \cr
         m &  0   &   0  &   0  &   0   &  -3/(2R)  &  \ldots \cr
         \vdots  &  \vdots &   \vdots  &   \vdots &
         \vdots &  \vdots  & \ddots \cr}~
\label{newnewnewMajmatrix}
\eeq
where we have {\it not}\/ performed any reordering of the rows and
columns corresponding to the excited Kaluza-Klein states.
If we truncate the Kaluza-Klein states at a chosen value $n_{\rm max}$,
we see that we always have an unpaired diagonal element of size
$(n_{\rm max}+\half)/R$.  Thus, when the excited Kaluza-Klein
states are integrated out, this leaves a net contribution to
the neutrino mass:
\beq
         m_\nu ~\approx~ {m^2 R \over n_{\rm max}+1/2}
               ~\approx~ {m^2 \over M_s}~.
\eeq
If we imagine
$m\lsim R^{-1}\approx 10^{-2}$ eV
(as would roughly be required to explain the neutrino oscillations)
and $M_s\approx 10$ TeV,
this gives rise to $m_\nu \approx 10^{-15}$ eV.
Thus, we see that sufficiently sizable neutrino oscillations
can be generated even with vanishingly small neutrino masses!

We conclude, then, that sizable neutrino oscillations can occur
 {\it regardless of the actual mass of the neutrino}, thanks to the
indirect masses and mixings of the Kaluza-Klein states.
Thus, if such a scenario can be realized within the context
of a fully realistic string model, then the recent
observations of neutrino oscillations can be re-interpreted
not as providing evidence for neutrino masses, but
rather as providing evidence for extra spacetime
dimensions!

\section{Conclusions, discussion, and open questions}
\setcounter{footnote}{0}

The scenarios that we have outlined in this paper
are certainly unorthodox, and so far they are only qualitative.
Certainly we have not performed a detailed comparison to
see if the wealth of existing experimental neutrino data
can be accommodated or explained in this manner.
Our goal, as we have stated throughout, has merely been to
provide a number of qualitative mechanisms which are capable of
yielding light neutrino masses without the {\it ad hoc}\/
introduction of heavy mass scales.  The important task of
implementing these mechanisms within self-consistent string
models remains.
It also remains necessary to perform a detailed comparison
with experimental data.

However, even at this preliminary stage, there are several
theoretical and phenomenological challenges that these scenarios face.
We would therefore like to conclude by discussing what some of
these challenges are.

One important theoretical issue which we have not addressed
concerns the dynamics of the branes to which the Standard-Model
fields
(but not the right-handed neutrino) are presumably restricted.
This issue may ultimately play an important
role in the seesaw mechanism because it has the potential to affect
the form of our mass mixing matrices.
Let us consider the matrix (\ref{newmatrix}) for concreteness.
As we discussed above,
the non-zero entries along the first row/column
reflect the coupling of the left-handed neutrino to the
excited Kaluza-Klein modes of the right-handed neutrino field.
Such couplings do not conserve momentum in the compactified
directions, but are allowed because the presence of the brane
to which the left-handed neutrinos are restricted breaks translational
invariance in these directions.

However, while these sorts of couplings are permitted in the
case of an infinitely rigid brane (as is typical in many
standard treatments), in reality the
brane can be expected to have a dynamics of its own.  In such
cases,
the couplings between the fields on the brane and the fields in
the bulk will become more complicated, and will presumably involve
the fluctuation modes of the brane itself.  A full analysis of
this question is beyond the scope of this paper.

Likewise, in the general matrix (\ref{newmatrix}),
we have set all remaining off-diagonal entries to zero.
As discussed in Sect.~2.2,
this reflects Kaluza-Klein momentum
conservation for couplings purely between
fields in the bulk.  However, this too is only an approximation:
in a complete theory (such as a string theory), we can expect
there to be higher-order couplings between different
Kaluza-Klein modes of the
bulk fields that arise indirectly through their
momentum-violating couplings to fields on
the brane.  However, once again this is a higher-order effect which
can be neglected at our level of approximation.  Furthermore,
even if such couplings are included,
we do not expect our primary results to be significantly affected.

Turning to phenomenological considerations,
we also find a number of outstanding questions.
One such question involves decays
such as $\pi\to \mu \bar\nu_\mu$.
Experimental measurements of the muon momentum spectrum reveal a very sharp
peak, confirming the fact that there is only one neutrino that
carries away momentum in such a decay.
However, in the higher-dimensional seesaw mechanisms we have been discussing,
the weak gauge charge of the gauge-eigenstate neutrino
is ultimately distributed over a whole tower of excited
Kaluza-Klein states.  Therefore, the infinite tower of Kaluza-Klein states
will,
in principle, partake in such processes.  Moreover, the lowest-lying
Kaluza-Klein
states can be relatively light.
In such situations, the muon momentum spectrum would therefore
be expected to be
quasi-continuous rather than discrete.
This therefore has the potential to severely constrain
our higher-dimensional seesaw scenarios (or generally, any scenario
in which the left-handed neutrino mixes with an infinite tower of Kaluza-Klein
neutrinos).  One redeeming feature of these scenarios, however, is the
extreme suppression of the Kaluza-Klein admixture.  For example,
in (\ref{eigenvec2}) we have seen that the Kaluza-Klein components are
suppressed by $mR$, where $mR\ll 1$.  Thus, these admixtures may
be sufficiently small to evade these sorts of experimental bounds.

Another important phenomenological issue concerns the ultimate stability
of the light (or vanishing) neutrino masses that are generated by
these higher-dimensional seesaw mechanisms.
Although these
mechanisms naturally yield light neutrino masses, these masses must still
be made stable against possible higher-order operators that can be generated
in the full effective (string) theory.
Discrete symmetries may be able to accomplish this,
but we have not investigated this possibility in this paper.
Nevertheless, it is interesting to note that any higher-order operators
that tend to generate effective heavy neutrino Majorana mass terms
will {\it not}\/ destabilize our results, for we
have already seen in Sect.~2.6 that the effect of the infinite
towers of Kaluza-Klein states is to eliminate the dependence on such
an external Majorana mass scale $M_0$ (regardless of its origin), automatically
replacing this scale with the new light scale
$\epsilon\equiv  M_0$ (modulo $R^{-1}$).
Thus, this feature may also provide some protection against higher-order
destabilizing effects.  Moreover, in the particular case of the Scherk-Schwarz
breaking of lepton number, we expect this breaking to vanish in the
$R\to\infty$
limit (with $M_s$ assumed fixed).  Therefore, any
such higher-dimensional operators must be suppressed by powers of $R$.

Thus, to summarize the main results of this paper,
we have seen that there exist several higher-dimensional analogues of
the usual seesaw mechanism
in which a mixing between a left-handed
neutrino and an infinite Kaluza-Klein tower of higher-dimensional
bulk fields
has the potential to
produce a light neutrino mass eigenstate
whose mass is suppressed
relative to the other mass scales in the problem.
Moreover, this occurs without the introduction of an
arbitrary high mass scale.
Even if such a bare Majorana mass
scale is present, we have seen that the higher-dimensional
seesaw mechanism with its summation over the Kaluza-Klein states
essentially eliminates this scale in favor of the large radius
of the extra spacetime dimensions.
We also pointed out a possible new mechanism, involving brane shifting
within Type~I string theory, for generating lepton-number
violation and thereby also inducing neutrino oscillations.
Finally, we also proposed
an explicit higher-dimensional seesaw mechanism in which
neutrino oscillations occur even without neutrino masses.
In this case, the neutrino oscillations occur
indirectly thanks to the masses and mixings of the Kaluza-Klein towers
of bulk neutrinos  or
through the usual flavor oscillations (which we have not discussed here).

Furthermore, as we remarked above, our higher-dimensional seesaw scenarios
are not restricted to the case of a single extra dimension.
Indeed, it is straightforward to extend these sorts of scenarios
to arbitrary numbers of extra dimensions, and similar results are obtained.
Moreover, one may even consider different radii (and even different fields)
for the different dimensions in order
to explain different types of neutrino oscillations.
This might therefore be capable of leading to a richer and more flexible
neutrino phenomenology
than is possible within the usual four-dimensional framework.

Thus, if these qualitative scenarios
can be made to operate within the context of fully realistic
models, they might provide higher-dimensional
mechanisms for generating light (or even vanishing) neutrino masses
without the need
for an intrinsic heavy mass scale for right-handed neutrinos.
However, as we have also seen, these sorts of scenarios face a number
of outstanding open questions.
It is important to stress that these problems are not specific
to the scenarios that we have put forth here, but rather generically
arise whenever the right-handed neutrino field is allowed to feel
large extra spacetime dimensions and whenever its
resulting Kaluza-Klein states can
also couple to the left-handed neutrinos.
Thus, the open questions  that we have discussed above have a generality
that transcends the specific mechanisms we have outlined, and will
ultimately need to be addressed in any scenario utilizing
higher-dimensional right-handed neutrinos.

\bigskip
\medskip
\leftline{\large\bf Acknowledgments}
\medskip

We wish to thank
K.~Benakli, B.~Campbell, C.~Johnson, J.~Lykken, R.~Myers, Y.~Nambu, J.~Pati,
A.~Pomarol, G.~Shiu, R.~Shrock, S.-H.H.~Tye, F.~Wilczek, and A.~Zaffaroni
for discussions.
We also wish to thank P.~Ramond for originally encouraging us
to consider the problem of neutrino masses in higher dimensions.

\bigskip
\medskip
\leftline{\large\bf Note Added}
\medskip

After this paper originally appeared in November 1998, some readers
apparently became confused regarding the violation of lepton number in
the higher-dimensional seesaw mechanism presented in Sect.~2.3.
We therefore reiterate that the kinetic-energy
terms in (\ref{tglagold}) {\it conserve}\/ lepton number, as required,
since the apparent non-conservation from the kinetic term
for the $N$ field
is precisely cancelled by the kinetic term for the $M$ field.
In other words, even though one Majorana component of each
Kaluza-Klein mass term has decoupled in (\ref{tglagold}),
the underlying Kaluza-Klein masses are indeed Dirac masses, as required.
It is only the {\it interaction}\/ between the left- and right-handed
neutrinos which breaks lepton number, as necessary in order
to obtain neutrino oscillations of the sort we have
been discussing.
Moreover, in order to avoid possible future confusion, we would like
to reiterate that our scenario of neutrino oscillations without neutrino
masses (as discussed in Sect.~4) has all of the following features:
the neutrino is exactly massless
for all values of $mR$, as shown in Fig.~\ref{orbfigurenew};
the massless eigenstate is unique and is primarily composed
of the gauge neutrino $\nu_L$ rather than the Kaluza-Klein modes,
as shown in (\ref{eigenvec2});
the resulting oscillations
are sizable and effectively
periodic, as shown in Fig.~\ref{oscfig2};
and this scenario can be realized directly in string theory through an orbifold
Scherk-Schwarz compactification, as discussed below (\ref{scherkschwarz}).
In such a compactification, the breaking of lepton number is topological
and fixed by the string boundary conditions.

\bigskip
\medskip

\bibliographystyle{unsrt}

\end{document}